\documentclass[aps,nofootinbib,showpacs,preprintnumbers,
]{revtex4} 

\usepackage{epsfig}
\usepackage{amssymb}
\usepackage{amsmath}

\begin{document}
\preprint{USM-TH-201, BI-TP 2006/44, hep-ph/0612130 v3}\footnote{
Version v3 as it appears in Phys.~Rev.~D {\bf 75}, 054016 (2007); changes made in v3
with respect to v2: Ref.~[25] is new; the ordering of the references and
grammatic and stylistic errors are corrected.} 
\title{Method for comparing finite temperature field theory results
with lattice data}

\author{G.~Cveti\v{c}}
  \email{gorazd.cvetic@usm.cl}
\affiliation{Center of Subatomic Studies}
\affiliation{Department of Physics, Universidad T\'ecnica
Federico Santa Mar\'{\i}a, Valpara\'{\i}so, Chile}
\author{R.~K\"ogerler}
 \email{koeg@physik.uni-bielefeld.de}
\affiliation{Department of Physics, Universit\"at Bielefeld, 
33501 Bielefeld, Germany}

\date{\today}

\begin{abstract}
The values of the presently available truncated 
perturbative expressions for the
pressure of the quark-gluon plasma at finite temperatures 
and finite chemical potential are trustworthy only at very large
energies. When used down to temperatures 
close to the critical one $(T_c)$, they suffer from
large uncertainties due to the renormalization scale freedom.
In order to reduce these uncertainties, we perform
resummations of the pressure  by applying
two specific Pad\'e-related approximants
to the available perturbation series for
the short-distance and for the long-distance contributions.
In the two contributions, we use two different 
renormalization scales which reflect different energy regions
contributing to the different parts.
Application of the obtained expressions at low temperatures
is made possible by
replacing the usual four-loop ${\overline {\rm MS}}$ beta function
for $\alpha_s$ by its Borel-Pad\'e resummation, 
eliminating thus the unphysical Landau singularities of $\alpha_s$.
The obtained results are remarkably insensitive to the chosen
renormalization scale and can be compared with lattice results --
for the pressure $p$, the chemical potential contribution $\Delta p$
to the pressure, and various susceptibilities. A good qualitative 
agreement with the lattice results is revealed down to temperatures 
close to $T_c$.
\end{abstract}

\maketitle

\section{Introduction}
\label{sec:introduction}
The behavior of QCD matter at nonvanishing temperature and (quark or
hadron) densities can be approached theoretically from two sides,
both technically and kinematically. On the one hand, QCD-lattice
calculations allow an analysis of the kinematic region of rather low
temperatures $T$ (around but above the phase-transition temperature
$T_c$, presently $T \alt 5 T_c$) and even lower values of the chemical
potentials $\mu_f$ ($f$ denotes the different quark flavors). 
The latter restriction is due to the fact that nonzero values
of the chemical potential $\mu$ render the weight function in the partition
function complex thus not permitting  direct application of the standard Monte
Carlo techniques. There are several tricks to 
circumvent this ``sign-problem'':
reweighting \cite{Fodor}, Taylor expansion around $\mu = 0$ \cite{Allton}, 
analytic continuation from imaginary $\mu$ values \cite{Forcrand},  canonical
formalism \cite{Krat}. All of them are trustworthy only for small
$\mu$ values.
On the other hand, the behavior at high $T$ and $\mu < 2 \pi T$ is
expected to be described reliably by finite temperature (and density) 
perturbation theory (FTPT). Significant progress has been made within this 
latter approach
during the last twenty years and the calculation of the thermodynamic
potential (free energy or, 
equivalently, the pressure function $p$) has recently
been pushed forward to the four-loop level, both for $\mu = 0$ \cite{KLRS1}
and for finite chemical potential \cite{Vuo}. This is
a big achievement because the corresponding truncated perturbation
series (TPS) is in powers of the QCD-strong coupling parameter $g$ rather
than $a \equiv g^2/(4 \pi^2) \equiv \alpha_s/\pi$, 
due to the well-known necessity of removing finite temperature 
infrared (IR) divergences by resumming the essential
IR-sensitive diagrams (``daisy diagrams'') to all orders. The final
series is up to 6th order in $g$, 
and that is essentially all that could be
expected from perturbation theory since terms proportional to $g^6$
(not $g^6~\ln~g$) and higher include genuine non-perturbative
contributions which could be accessed only by non-perturbative methods,
e.g. lattice calculations. 

Of course, any perturbatively obtained 
result is expected to represent the true physical situation only when the
coupling parameter is small enough, which is the case in QCD at
sufficiently high temperatures. Nevertheless,
evaluations of the 6th order result (for $p$ and for 
$\Delta p = p(\mu) - p(0)$)
down to temperatures below 1 GeV have been made in the literature, and
the results could be brought in accordance with 
non-perturbative lattice data (available for $T$ up to $1$ GeV). 
At first sight this could be considered a triumph of finite 
temperature perturbation theory, and might tentatively be attributed
to the higher order available. A second and more careful glimpse, however,
reveals several sobering observations. 
The first problem is that the convergence behavior of the truncated 
perturbation series is manifestly weak: 
if increasingly higher orders in the series are
added, the corresponding partial sums are changing wildly, jumping up and
down. Only the step from 5th to 6th order has signs of
moderation, but it is by no means clear whether this happens by chance
or indicates a systematic improvement of the situation. 
Further, a common relatively high 
renormalization scale (RScl)\footnote{In our calculations we will denote the
renormalization scale as $\nu$, keeping the symbol 
$\mu$ for the chemical potential.
The reader should be aware of the fact that in Refs.~\cite{KLRS1} and 
\cite{Vuo}
the common RScl (in ${\overline {\rm MS}}$ scheme) 
is denoted by ${\overline \mu}$ and $\bar \Lambda$, respectively.}  
$\nu \approx 2 \pi T$ ($ \gg T$) is used in such evaluations.
For comparison with independent lattice data, the TPS
results are used down to small $T$ values
where the lattice results are available.
But here, another problem occurs: in each available order the
corresponding TPS shows a dependence of the chosen RScl
$\nu$, which is particularly strong for lower temperatures,
thus making a consistent comparison of perturbative results (when
used at such low energies) with lattice data doubtful. In fact
we have no clear physical motivation for the best choice of $\nu$. In
general, $\nu$ should be chosen such that large (momentum dependent) 
logarithms in the TPS coefficients are avoided, which means -- within an
asymptotically free theory like QCD -- that one is always on the safe side if
$\nu$ is taken to be near the lowest energy scales involved in the
considered quantity. For the quark-gluon plasma at (high) temperature
$T$ (and for given chemical potentials $\mu_f$), what is the appropriate
energy scale? Usually in the literature, the energy $2 \pi T$ of the
lowest nonzero mode is taken as a 
measure since it determines the average energy
of the constituents. But due to the collective effects additional (lower)
physical scales are generated, namely the electric and magnetic screening
masses (being of order $gT$ and $g^2T$, respectively). So what to choose
for $\nu$? This question is not a purely theoretical one, but of considerable
practical importance, because of the mentioned strong dependence of the TPS
on $\nu$. And even if we neglect this intertwining of different scales
within the thermodynamic quantities, and stick to $2 \pi T$ as the relevant
scale, we observe that a change of $\nu$ by a single factor of 2
implies such a wide variation of the perturbative results at $T \sim T_c$
that no firm conclusion can be drawn as to the matching with lattice
data (cf.~Figs.~\ref{FigpvsTRScl} and \ref{FigDpvsTRScl}). 
It is clear, therefore, that no successful matching
procedure can be obtained unless the strong RScl dependence is pinched 
down by improving the perturbative results.

Within the present paper we offer a way for avoiding this unwanted 
(and unphysical) ambiguity  by applying perturbation-theory-improving 
resummations of the basic TPS's. In this way we obtain what
we consider a more reliable, but still perturbation-theory-based, description
of the interesting thermodynamic quantities (here the pressure $p$) which,
among other things, allow for a more credible 
comparison with lattice results. The
method rests on replacing the (partially resummed) TPS
by approximants which are much more stable
under the variation of $\nu$ than the TPS's themselves. 
It is well known that specifically Pad\'e
approximants \cite{Padebook} for 
physical (measurable) quantities
are stable under $\nu$ variation \cite{Gardi}, whereas
similar improved Baker-Gammel approximants  show this
stability exactly \cite{Cvetic}.\footnote{
In addition, an extension of such RScl-independent approximants can
be constructed, giving results which are simultaneously RScl- and
scheme-independent \cite{CK2}. In the present work, the ${\overline {\rm MS}}$
scheme is being used throughout.}
In a recent work \cite{GRp} we have 
utilized Pad\'e-related approximants to produce RScl-stable expressions for the 
pressure of the quark-gluon plasma at finite (large)
temperature and zero chemical potential. Here, we apply the same 
technique to the case of finite nonzero quark chemical potential. In
both cases, the gratifying fact is the high order of the available TPS
which allows the use of higher-order Pad\'e-related approximants and the
choice of the most appropriate ones (see later).

{}From the technical point of view we encounter a specific problem
when applying Pad\'e-related approximants directly to finite temperature
perturbation theory. This is due to the fact that two different 
(infinite) classes of diagrams get involved in the whole mechanism:
on the one hand those whose resummation is necessary for taming the
finite-temperature IR divergences and which lead to 
contributions in powers of $g$ (the so-called ``daisy diagrams'');
on the other hand the other diagrams, which give contributions
in powers of $g^2$ and whose conversion from the TPS (polynomials) 
into Pad\'e approximants (rational functions) had been known
to result in RScl stability (which is exact in the large-$\beta_0$ limit
in the case of the diagonal Pad\'e approximants \cite{Gardi}).
Therefore, care
has to be taken to avoid double counting when performing both resummations.
As we have shown in Refs.~\cite{GRp}, a safe method consists in 
first decomposing 
the pressure into two parts,
one containing the low-energy (effectively zero) modes and being
responsible for the long-distance behavior of the correlation
functions, and the other stemming from the nonzero modes and 
determining the short-range physics. Since both regions are,
in principle, 
separately accessible by experiments, the corresponding expressions are
physical in the sense that they should be independent of the 
renormalization scale $\nu$ (at least up to the specified order
in $g$). Further, no (infinite) resummation enters into
the perturbation expressions of those parts when dimensional reduction
is applied \cite{Gross}. Therefore, one can safely apply 
Pad\'e-related resummation
to both parts independently and thus obtain two expressions which are
both almost independent of the RScl choice. When using them
down to low energies $(T \sim T_c)$, we face another problem: the
coupling parameter $a(Q^2)$ acquires unphysical (Landau)singularities
at low energies $Q \alt 1$ GeV if using ${\overline {\rm MS}}$ TPS $\beta$ 
function. We circumvent this problem by using again
appropriately resummed versions of $\beta$. We finally
end up with expressions for the pressure and for other measurable
quantities (quark number susceptibilities) which are almost free of
the RScl uncertainty and therefore apt for comparison with lattice data.

In Sec.~II we present the perturbative results on which our analysis rests
and describe how to perform the physical separation of the pressure into
the long-range and the short-range part. 
Section III contains an analysis of the
possible resummation procedures which leads to the optimal choice. We then
present the numerical results, 
where we put the main emphasis both on the effect of finite $\mu$ values 
and on the comparison with the corresponding lattice data.
Appendix \ref{app1} compiles basic formulae and expressions, available in the
literature and adapted to our approach, including expressions for the
coefficients of the TPS's. Appendix \ref{app2} describes the method which allows
us to extrapolate the QCD renormalization group equation to sufficiently
low energies, thereby circumventing the unphysical Landau singularities of
the coupling parameter.

\section{Separation of Long- and Short-Distance Pressure}
\label{sec:PEs}

Our starting point are the FTPT results
for the pressure of the quark-gluon plasma which, for a homogeneous
system, is equal to the free energy per volume (up to a sign difference)
-- considered as a function of its temperature $T$ and the chemical 
potentials of the various quark flavors $\mu_f~(f=1,...n)$. It has
been calculated up to ${\cal O} (g^6 \ln g)$ 
by K.~Kajantie {\it et al.\/}~\cite{KLRS1}
for the case of vanishing chemical potentials, and for the general case
$(\mu_f \not= 0)$ by A.~Vuorinen \cite{Vuo}. These high order 
results include the summation of an infinite class of certain diagrams
-- necessary for taming an IR singularity which occurs only at
$T \not= 0$. The final results could be practically achieved only by a
technical trick, namely by separating the energy-momentum region of the
contributing modes into three parts, characterized by the momentum scales
$2 \pi T,~~gT$ and $g^2T$, such that the full pressure $p$ is decomposed 
according to 
\begin{equation}
p = p_{\rm E} + p_{\rm M} + p_{\rm G} \ .
\label{psum}
\end{equation}
Note that this decomposition makes strict sense only for high enough
temperatures where $g (T) \alt 1$. Here, $p_{\rm E}$
represents the contributions of all degrees of freedom associated with
the nonvanishing Matsubara modes, whereas $p_{\rm M} + p_{\rm G}$ comprises the 
contributions of the zero modes (of bosonic fields), thereby implicitly 
representing also the necessary sum over all (daisy) diagrams. The
latter ones are static modes, hence their contributions can be 
effectively described by a three-dimensional (in general $d$-dimensional)
purely bosonic field theory (dimensional reduction \cite{Gross}) determined 
by the electrostatic QCD (EQCD) Lagrangian
\begin{equation}
{\cal L}_{\rm EQCD} = \frac{1}{2} {\rm Tr} \tilde{F}^2_{ij} 
+ {\rm Tr} [D_i, \tilde{A}_0]^2 
+ m_{\rm E}^2 {\rm Tr} \tilde{A}^2_0 
+ \lambda^{(1)}_{\rm E} ({\rm Tr} \tilde{A}^2_0)^2 
+ \lambda^{(2)}_{\rm E} {\rm Tr} \tilde{A}_0^4 
+ i \frac{g^3}{3 \pi^2} \left( \sum_f \mu_f \right)
{\rm Tr} \tilde{A}^3_0 + ...
\label{EQCD}
\end{equation}
Here $\tilde{A}_0$ denotes an effective ($d$-dimensional) scalar field and the
$\tilde{A}_i~(i = 1, ...d)$ define a $d$-dimensional vector field, both in matrix
notation $(\tilde{A}_\mu \equiv \tilde{A}_\mu^a T^a)$;\footnote{
The fields $\tilde{A}^a_0,
\tilde{A}^a_i$ entering here are not identical to the gluon fields in 
${\cal L}_{\rm QCD}$ but are the effective fields obtained after the high-energy
modes have been integrated out.}
$D_i = \partial_i - i g_{\rm E}  \tilde{A}_i$; 
$\tilde{F}_{ij} = (i/g_{\rm E} ) [D_i, D_j]$.
The parameters of this effective theory are the (electrostatic) screening
mass $m_{\rm E}~(\sim gT)$, the effective coupling parameter $g_{\rm E} ^2~(\sim g^2T)$
and the four-vertex couplings $\lambda^{(1)}_{\rm E}, \lambda^{(2)}_{\rm E} (\sim g^4 T)$.
In the case of $d=3,~~\lambda^{(1)}_{\rm E}$ and $\lambda^{(2)}_{\rm E}$ are not
independent and one can choose $\lambda^{(2)}_{\rm E} = 0$ -- this will be
done in the following. There are additional coupling parameters
connected with Lagrangian operators of higher dimensions, since
${\cal L}_{\rm EQCD}$ defines a non-renormalizable theory which makes sense
only for momenta below a certain (UV)cutoff $\Lambda_{\rm E}$. In our
case $\Lambda_{\rm E}$ separates the region of momenta $\sim 2 \pi T$ from
the momenta $\sim gT$ and smaller.  

The effective parameters can be connected to the parameters of the
underlying QCD by means of the well-known matching 
procedure \cite{Braaten:1996ju} 
yielding
\begin{eqnarray}
m_{\rm E}^2 & = & T^2 {\Bigg \{} g^2 \left[ 
A_4 +  \epsilon \left( A_5^{(\nu)} \ln \frac{\nu_c}{2 \pi T}
+ A_5 \right) + {\cal O}(\epsilon^2) \right] 
\nonumber \\
&& 
+ \frac{1}{(4 \pi)^2} g^4 \left[
A_6^{(\nu)} \ln \frac{\nu_c}{2 \pi T} 
+ A_6 + {\cal O} (\epsilon) \right] + 
{\cal O} (g^6) {\Bigg \}} \ , 
\label{mE2o1}
\\
g_{\rm E} ^2 & = & T {\Bigg \{} g^2 + \frac{1}{(4 \pi)^2} g^4 \left[
A^{(\nu)}_7 \ln \frac{\nu_c}{2 \pi T} + A_7 + {\cal O} (\epsilon)
\right] 
+ {\cal O} (g^6)  {\Bigg \}} \ , 
\label{gE2o1}
\\
\lambda^{(1)}_{\rm E} & = & T {\Bigg \{} 
\frac{1}{(4 \pi)^2} g^4 \left[ 
\beta_{\rm E4} + {\cal O}(\epsilon) \right]
+ {\cal O} (g^6) {\Bigg \}}, \qquad  \lambda_{\rm E}^{(2)} = 0 \ .
\label{lE12}
\end{eqnarray}
Here, $\epsilon = (3-d)/2$; 
coefficients $A_4$-$A_7$, $A_5^{(\nu)}$-$A_7^{(\nu)}$ and
$\beta_{\rm E4}$ are complicated functions of the chemical potentials,
the latter appearing in the coefficients via the dimensionless
quantities ${\overline \mu_f} \equiv \mu_f/(2 \pi T)$. 
Their expressions are collected in Appendix \ref{app1}, together with other 
coefficients to appear in Eqs.~(\ref{pEo1})-(\ref{pGo1}).
Note that the common RScl $\nu_c$ appears in these 
expressions. The effective mass $m_{\rm E} \sim gT$ arises due to
the color-electric screening.

Since there is, in addition, color-magnetic screening at energies proportional
to the corresponding magnetic screening mass 
$m_{\rm M} \sim g^2 T$, the long-distance part of the pressure can be further
subdivided  into $p_{\rm M}$ and $p_{\rm G}$, where $p_{\rm M}$ is determined by 
${\cal L}_{\rm EQCD}$ and $p_{\rm G}$ by the (magnetostatic) Lagrangian
\begin{equation}
{\cal L}_{\rm MQCD} = \frac{1}{2} {\rm Tr} \tilde{F}_{ij} + ...
\label{MQCD}
\end{equation}
with
$\tilde{F}_{ij} = (i/g_{\rm M}) [\tilde{D}_i, \tilde{D}_j]$ and
$\tilde{D}_i = \partial_i - ig_{\rm M} \tilde{A}_i$.
This Lagrangian defines the effective theory for energies below
$\Lambda_{\rm M}$, i.e., for energies $\sim g^2 T$ and smaller.
A similar matching procedure as before determines $g_{\rm M}$ in terms of the 
parameters of the higher-energy Lagrangian ${\cal L}_{\rm EQCD}$ and gives
\begin{equation}
g^2_{\rm M} = g_{\rm E} ^2 + {\cal O} (g^3).
\label{gM2}
\end{equation}

In the case of nonzero chemical potentials, 
two scales get involved ($T$ and $\mu \equiv \mu_f$) and, therefore,
the concept of dimensional reduction is expected to be applicable
only if the magnitude of the chemical potentials is small compared
to $2 \pi T$ \cite{HLP}. From comparison with numerical results for
correlation lengths, it is expected that the restriction $\mu
\le 4 T$ is safe.

Based on Lagrangians (\ref{EQCD}) and (\ref{MQCD}), 
and on the ordinary QCD-Lagrangian,
the various parts  of the pressure have
been calculated perturbatively by Vuorinen \cite{Vuo}. The
calculations are based on dimensional regularization with a common 
RScl $\nu_c$ (the notation $\bar \Lambda$ is used in Ref.~\cite{Vuo}
for the common RScl).

The result for $p_{\rm E}$ is
\begin{eqnarray}
\frac{p_{\rm E}}{T} &= T^3 & {\Bigg [} 
A_1 + g^2 \left( A_2 + {\cal O} (\epsilon) \right)
\nonumber \\
&& + \frac{1}{(4 \pi)^2} g^4 \left(
\frac{1}{\epsilon} 6 A_4 +  A_3^{(\nu)}
\ln \frac{\nu_c}{2 \pi T} + A_3 + {\cal O} (\epsilon)\right) 
\nonumber \\
&& 
+ \frac{1}{(4 \pi)^4} g^6 \left( \beta_{\rm E1} + {\cal O} (\epsilon) \right) 
+ {\cal O} (g^8)
{\Bigg ]} \ .
\label{pEo1}
\end{eqnarray}
Coefficients $A_i~(i=1,2,3)$ and $A_3^{(\nu)}$ are collected in 
Appendix \ref{app1}. Coefficient $\beta_{\rm E1}$ at $g^6$ is still unknown.
However, $\beta_{\rm E1}$ must include a term proportional to $1/\epsilon$.
The $1/\epsilon$-terms will be disposed of in the following 
because such terms must cancel in the sum
(\ref{psum}), and the finite part of $\beta_{\rm E1}$
will contain a free (adjustable) parameter later in this work.

The results for $p_{\rm M}$ and $p_{\rm G}$, which can be obtained from 
the effective Lagrangians (\ref{EQCD}) and (\ref{MQCD}), 
respectively, are \cite{Vuo}
\begin{eqnarray}
\frac{p_{\rm M}}{T}  & = & 
m_{\rm E}^3 \frac{2}{3 \pi} \left\{ 1 + \frac{1}{4 \pi}
3^2 \frac{g_{\rm E} ^2}{m_{\rm E}} \left[ - \frac{3}{4} - \ln \frac{\nu_c}{2 m_{\rm E}}
\right] \right.
\nonumber \\
& + & \frac{1}{(4 \pi)^2} 3^3 \left( \frac{g_{\rm E} ^2}{m_{\rm E}} \right)^2
\left[ - \frac{89}{24} - \frac{\pi^2}{6} + \frac{11}{6} \ln 2 \right] 
\nonumber \\
& + & \frac{1}{(4 \pi)^3} 3^4 \left( \frac{g_{\rm E} ^2}{m_{\rm E}} \right)^3
\left[ \alpha_{\rm M1} 8 \ln \frac{\nu_c}{2 m_{\rm E}} + \beta_{\rm M1} +
\frac{5}{81}  (\sum_f {\overline \mu_f})^2
\left( \alpha_{\rm M2} \ln \frac{\nu_c}{2 m_{\rm E}}
+ \frac{1}{4} \beta_{\rm M2}\right) \right] \nonumber \\
& - & \left. \frac{15}{8 \pi} \frac{\lambda_{\rm E}^{(1)}}{ m_{\rm E}} \right\} + 
\nonumber \\
& + & \frac{1}{\epsilon}  m_{\rm E}^3 \frac{2}{3 \pi} \left\{ \frac{1}{4 \pi}
3^2 \frac{g_{\rm E} ^2}{ m_{\rm E}} \left( - \frac{1}{4} \right) + 
\frac{1}{(4 \pi)^3} 3^4 \left( \frac{g_{\rm E} ^2}{ m_{\rm E}}\right)^3
\left[ \alpha_{\rm M1} + \frac{5}{81}
\alpha_{\rm M2} \frac{1}{4} ( \sum_f {\overline \mu_f} )^2 \right] \right\}
\nonumber\\
& + & {\cal O}(\epsilon) + {\cal O}(g_{\rm E} ^8/ m_{\rm E}) \ ,
\label{pMo1}
\end{eqnarray}
\begin{eqnarray}
\frac{p_{\rm G}}{T}  & = & \frac{2}{3 \pi} \frac{1}{(4 \pi)^3} 3^4 
\left( g^2_{\rm M}\right)^3 \left[ \alpha_{\rm G} 8 \ln \frac{\nu_c}{2m_{\rm M}} + 
\delta_{\rm G}  + {\cal O}(\epsilon) \right] + \hspace{5.5cm} \nonumber \\
& + & \frac{1}{\epsilon} \frac{2}{3 \pi} \frac{1}{(4 \pi)^3} 3^4 
\left( g^2_{\rm M} \right)^3 \alpha_{\rm G} \ .
\label{pGo1}
\end{eqnarray}
Here, $m_{\rm M}$ is the magnetic screening mass $m_{\rm M} = 3 g_{\rm M}^2$ ($\sim g^2 T$),
and the notation ${\overline \mu_f} \equiv \mu_f/(2 \pi T)$ is used.
We observe that the chemical potentials show up both explicitly via ${\overline \mu_f}$
and implicitly via $g_{\rm E} $ and $ m_{\rm E}$. 
Coefficients $\alpha_{\rm M1}, \alpha_{\rm M2}, \alpha_{\rm G},
\beta_{\rm M1}, \beta_{\rm M2}$ are independent  of $\mu_f$ (see Appendix \ref{app1}),
and only $\beta_{\rm M2}$ is not yet known.
In Eqs.~(\ref{pMo1}) and (\ref{pGo1}) 
we have already separated all ``divergent'' terms 
(proportional to $1/\epsilon$) from the finite contributions.
The quantity $p_{\rm M}/T$ starts effectively with order $g^3$ and
$p_{\rm G}/T$  with order $g^6$. The full $\sim g^6$ term in
$p_{\rm G}/T$ cannot be determined in a perturbative way;
however, $\delta_{\rm G}$ was estimated in Ref.~\cite{DiRenzo:2006nh} to
be $\delta_{\rm G} = -0.2 \pm 0.6$. Here, it will be treated as a
free parameter within the limits $\delta_{\rm G} = 0 \pm 1$.
On the left-hand sides of Eqs.~(\ref{pEo1})-(\ref{pGo1}),
the common denominator $T$ must in fact be replaced by
$T \nu_c^{-2 \epsilon}$; however, the common factor
$\nu_c^{-2 \epsilon}$ reduces to unity in the 
$\epsilon \to 0$ limit, and will be ignored because the
sum $p_{\rm E} + p_{\rm G} + p_{\rm M}$ is finite in this limit.

Our main task now consists in deducing from these formulae expressions for
the physical short-range and long-range parts of the pressure. Thereby
the word ``physical'' indicates that they lead to measurable effects
(for instance, the long and/or short-range behavior of static correlation
functions, etc.).
We proceed in several steps (see also Ref.~\cite{GRp}):
\begin{itemize}
\item[(i)] Regularization:\hfill\break
The long-range part of the pressure is represented by
$p_{\rm M} + p_{\rm G} \equiv p_{\rm M+G}$ 
(both are due to zero mode contributions). We regularize
$p_{\rm M+G}$ by adding to $p_{\rm M+G}/T$ the following counterterm:
\begin{equation}
CT = \frac{1}{\epsilon} 6 \left\{ \frac{1}{(4 \pi)^2} g_{\rm E} ^2  m_{\rm E}^2 -
36 \frac{1}{(4 \pi)^4} \left[ g^6_{\rm E} \left( \alpha_{\rm M1} +
\alpha_{\rm M2} \frac{5}{324}
(\sum_f {\overline \mu_f})^2 \right) + g^6_{\rm M} \alpha_{\rm G} \right] \right\}.
\label{CT}
\end{equation}
The same counterterm has to be subtracted from $p_{\rm E}/T$. By
expanding the effective-theory-parameters $g_{\rm E} $ and $ m_{\rm E}$ in powers of
$g$ [cf.~Eqs.~(\ref{mE2o1}), (\ref{gE2o1})], 
one can explicitly show that in this way the
$1/\epsilon$-term of order $g^4$ in expression (\ref{pEo1})
for $p_{\rm E}/T$ gets cancelled. Further,
the $1/\epsilon$-term of order $g^6$ included in the otherwise unknown
$\beta_{\rm E1}$ coefficient must also get cancelled when counterterm (\ref{CT}) is
subtracted from $p_{\rm E}/T$ of Eq.~(\ref{pEo1}). 
Counterterm (\ref{CT}) contains one finite term of 
${\cal O} (g^4)$ stemming from $(1/\epsilon) g_{\rm E} ^2  m_{\rm E}^2$ since $ m_{\rm E}^2$
includes a term proportional to $g^2 \epsilon$, Eq.~(\ref{mE2o1}). 
This finite term then shows up in the new subtracted expression for $p_{\rm E}/T$.
In addition, counterterm (\ref{CT}) contains several finite terms of
${\cal O} (g^6)$ which show up in the new subtracted expression for $p_{\rm E}/T$.
Finally, the limit
$\epsilon \to 0$ can be performed, yielding finite results both for 
$p_{\rm E}$ and for $p_{\rm M+G}$.
\item[(ii)] Reconstruction of the factorization scale and introduction
of various RScl's:\hfill\break
We have already noted that a common renormalization scale (denoted here
as $\nu_c$) has been used for the perturbative calculations leading
to Eqs.~(\ref{pEo1})-(\ref{pGo1}). On the other hand, when constructing the 
physical long- and short-range contributions, respectively, we decompose
the whole energy range in the way addressed before, namely
\begin{equation}
{\cal O}(g^2T) < \Lambda_{\rm M} < {\cal O}(gT) < \Lambda_{\rm E} < {\cal O} (2 \pi T) \ .
\end{equation}
Therefore, these factorization scales $\Lambda_{\rm E}$ and $\Lambda_{\rm M}$ which
separate the different energy regions have to emerge in the physical
expressions -- but only in such a way that, when adding all three
contributions they completely disappear. 
We bring the $\Lambda_{\rm E}$ factorization scale to light in the following way:
the scale $\nu_c$ in expressions (\ref{pEo1})-(\ref{pGo1}),
where $p_{\rm E}/T$ and $p_{\rm M+G}/T$ are modified in the afore-mentioned way
by the counterterm (\ref{CT}), is
interpreted simultaneously as the factorization scale $\Lambda_{\rm E}$ and
as the common RScl $\nu_c$. We then evolve $g(\nu=\Lambda_{\rm E})$ in
$p_{\rm E}$ to $g(\nu_{\rm E})$ where $\nu=\nu_{\rm E}$ is a new, physically
more adequate higher RScl: $\nu_{\rm E} \sim 2 \pi T$. On the other hand,
in $p_{\rm M+G}$, we evolve $g(\nu=\Lambda_{\rm E})$, which
appears implicitly there (explicitly in the RScl-independent
$g_{\rm E} ,  m_{\rm E}, \lambda_{\rm E}^{(1)}$),
to $g(\nu_{\rm M})$ where $\nu=\nu_{\rm M}$ is a new, physically more
adequate lower RScl: $\nu_{\rm M} \sim  m_{\rm E}$ ($\sim g T$). The evolution
is performed according to perturbative renormalization group
equation (RGE), requiring RScl independence of $p_{\rm E}$, on the
one hand, and of $g_{\rm E} ^2$ and $ m_{\rm E}^2$ (and thus of $p_{\rm M+G}$),
on the other hand, since all these quantities are physical.
In $p_{\rm E}$ this results in $\ln \nu_{\rm E}$-dependent terms in 
the coefficients of TPS, and in $g_{\rm E} ^2$ and $ m_{\rm E}^2$ (which enter
$p_{\rm M+G}$) this results in $\ln \nu_{\rm M}$-dependent terms
in the coefficients of their TPS's.
The coefficients of expansion of $p_{\rm E}$ in powers of $g^2(\nu_{\rm E})$
then have explicitly the ``genuinely'' $\Lambda_{\rm E}$-dependent parts,
and RScl-dependent parts. The coefficients of expansion of
$p_{\rm M+G}$ in powers of $(g_{\rm E} ^2/ m_{\rm E})$ remain unchanged,
with $\Lambda_{\rm E}$ dependence as before, 
while the coefficients of expansion of $ m_{\rm E}^2$ and
$g_{\rm E} ^2$ in powers of $g^2(\nu_{\rm M})$ obtain RScl dependence
and, at the order considered, lose $\Lambda_{\rm E}$ dependence.\footnote{
This is not so in the scalar $g^2 \phi^4$ theory, where the
requirement of RScl independence of the Debye screening mass
yields, at ${\cal O}(g^4)$, a residual dependence on the
factorization scale (cf.~Ref.~\cite{GRp}, second entry).}
Formally, the two RScl's $\nu_{\rm E}$ and $\nu_{\rm M}$ can take on
arbitrary values in these expressions.

\item[(iii)] Determination of the $\Lambda_{\rm E}$-dependent part of 
coefficient at ${\cal O}(g^6)$ in $p_{\rm E}$:
\hfill\break
The sum $p_{\rm E} + p_{\rm M+G}$ has to be independent
of the factorization scale $\Lambda_{\rm E}$. It can be checked explicitly
that this independence is true up to ${\cal O}(g^4)$.
Although the coefficient at $g^6$ in $p_{\rm E}$ is not known,
its $\Lambda_{\rm E}$ dependence is dictated by the condition of
$\Lambda_{\rm E}$ independence of $p_{\rm E}+p_{\rm M+G}$ at ${\cal O}(g^6)$,
when the latter quantity is expanded in powers of a common
$g = g(\nu)$.
Note that $\Lambda_{\rm E}$ dependence of $p_{\rm M+G}$ at ${\cal O}(g^6)$
is known, cf.~Eqs.~(\ref{pMo1})-(\ref{pGo1}).
Further, $\nu$ dependence of the coefficient at $g^6$ in
$p_{\rm E}$ is known from the requirement of the RScl independence of $p_{\rm E}$.
The remaining unknown part of the coefficient at ${\cal O}(g^6)$ 
in $p_{\rm E}$ (we will call it $\delta_{\rm E}$) 
is then independent of $\Lambda_{\rm E}$ and of $\nu$,
and can again be freely adjusted.
\end{itemize}
Performing these steps we finally obtain the following form of the
physical decomposition of the pressure function into short- and long-distance
parts (we denote these physical parts by a bar)
\begin{equation}
p = {\bar p}_{\rm E} + {\bar p}_{\rm M+G}.
\end{equation}
The long-distance part, representing the contributions of momenta below the 
factorization  scale $\Lambda_{\rm E}$ (when $gT < \Lambda_{\rm E} < 2 \pi T$) takes
the form
\begin{eqnarray}
\frac{1}{T} {\bar p}_{\rm M+G} & = & \frac{2}{3 \pi}  m_{\rm E}^3 \left\{ 1 + 
\frac{1}{4 \pi} 3^2 \left( \frac{g_{\rm E} ^2}{ m_{\rm E}} \right) \left[ - \frac{3}{4}
- \ln \frac{\Lambda_{\rm E}}{2 m_{\rm E}} \right] \right.  
\nonumber \\
& + & \frac{1}{(4 \pi)^2} 3^3 \left( \frac{g_{\rm E} ^2}{ m_{\rm E}} \right)^2 
\left[ - \frac{89}{24} - \frac{\pi^2}{6} + \frac{11}{6} \ln 2 \right] 
\nonumber \\
& + & \frac{1}{(4 \pi)^3}  3^4 \left( \frac{g_{\rm E} ^2}{ m_{\rm E}} \right)^3 
\left[ 8 \alpha_{\rm M1} \ln \frac{\Lambda_{\rm E}}{2 m_{\rm E}} + 8 \alpha_{\rm G} \ln 
\frac{\Lambda_{\rm E}}{6 g_{\rm E} ^2} + \right. \nonumber \\ 
& + & \left. \beta_{\rm M1} + \delta_{\rm G} - \frac{20}{3^5} n^2_f {\widetilde \mu}_1^2 
\left(\ln
\frac{\Lambda_{\rm E}}{2 m_{\rm E}} - \frac{3}{16} \beta_{\rm M2} \right) \right] 
\nonumber \\
&& \left. - \frac{15}{8 \pi} \frac{\lambda_{\rm E}^{(1)}}{ m_{\rm E}} \right\} \ ,
\label{pMGeff}
\end{eqnarray}
where we used notation (\ref{not1})-(\ref{not2}) for the
chemical potential parameter ${\widetilde \mu}_1$.
The parameters $ m_{\rm E}, g_{\rm E} $ and $\lambda_{\rm E}^{(1)}$ of the effective theory
EQCD are defined by their expansion into powers of $g (\nu) \equiv g$
\begin{eqnarray}
 m_{\rm E}^2 & = & T^2 A_4 g^2 \left\{ 1 + \left( \frac{g}{2 \pi} \right)^2
\left[ 2 \beta_0 \ln \left( \frac{{\nu}}{2 \pi T} \right) + \frac{1}{4} \frac{A_6}{A_4} \right]
+ {\cal O} \left( \left( \frac{g}{2 \pi} \right)^4 \right) \right\} 
\label{mE2}
\\
g_{\rm E} ^2 & = & T g^2 \left\{ 1 + \left( \frac{g}{2 \pi} \right)^2 
\left[ 2 \beta_0 \ln \left( \frac{{\nu}}{2 \pi T} \right) + \frac{1}{4} A_7 \right] + 
{\cal O} \left( \left( \frac{g}{2 \pi} \right)^4 \right) \right\} 
\label{gE2}
\\
\lambda_{\rm E}^{(1)} & = & T g^4 \; \frac{2}{3} \frac{1}{(4 \pi)^2} (9 - n_f)
\left\{ 1 + {\cal O} \left( \left( \frac{g}{2 \pi} \right)^2 \right) \right\}.
\label{l1}
\end{eqnarray}
Here, $\beta_0 = (1/4) (11 - 2 n_f/3)$ is the one-loop QCD 
RGE coefficient, $n_f$ being the number of active quark flavors.
The RScl $\nu$ in Eqs.~(\ref{mE2})-(\ref{l1}) 
will be $\nu = \nu_{\rm M}$ ($\sim  m_{\rm E} \sim g T$), 
and we fix it according to relation (\ref{RScls}).
Note that, by the afore-described procedure,
the expansion coefficients of $ m_{\rm E}^2$ and $g_{\rm E} ^2$
at the considered order
do not have any dependence of the factorization scale $\Lambda_{\rm E}$,
but are RScl-dependent.
The last term in expansion (\ref{pMGeff}) involves
the third EQCD matching parameter $\lambda^{(1)}_{\rm E}$ 
\cite{Nadkarni:1988fh,Landsman:1989be}, which is
independent of the chemical potentials $\mu_f$. 
This term can be expressed as a power series in 
powers of $g$, but only the
leading term is known $\lambda^{(1)}_{\rm E}/ m_{\rm E} \propto g^3$.
We prefer to express it in powers of the
EQCD parameter $g_{\rm E} ^2/ m_{\rm E}$ (Ref.~\cite{GRp})
\begin{equation}
\lambda^{(1)}_{\rm E} = \frac{2}{3} \frac{1}{(4 \pi)^2} (9 - n_f) 
A_4 \;  m_{\rm E} \left( \frac{g_{\rm E} ^2}{ m_{\rm E}} \right)^3 
\left[ 1 + {\cal O} \left( \left( \frac{g_{\rm E} ^2}{ m_{\rm E}} \right)^2 \right) \right]
\ .
\label{l1b}
\end{equation}
While coefficients $\alpha_{\rm M1}, \alpha_{\rm G}, \beta_{\rm M1}$ 
in Eq.~(\ref{pMGeff}) and
$A_i~(i = 4,...7)$ in Eqs.~(\ref{mE2})-(\ref{gE2}) 
are known and collected in Appendix \ref{app1}, 
parameter $\delta_{\rm G}$ is well estimated \cite{DiRenzo:2006nh},
and $\beta_{\rm M2}$ is unknown. The dependence on the chemical
potentials $\mu_f (f = 1, ..., n)$ in ${\bar p}_{\rm M+G}$
appears explicitly in the term
proportional to ${\widetilde \mu}_1^2$ and
implicitly via the parameters $ m_{\rm E}, g_{\rm E} $ which contain
$\mu_f$-dependent coefficients $A_i$.
The RScl $\nu=\nu_{\rm M}$ appears in ${\bar p}_{\rm M+G}$ 
implicitly, via expansions (\ref{mE2}) and (\ref{gE2}) for
$ m_{\rm E}^2$ and $g_{\rm E} ^2$.

The physical short-distance part, determined by the energy-momentum range
above the factorization scale $\Lambda_{\rm E}$, can be written in a dimensionless
form as follows:
\begin{equation}
\frac{1}{T^4} {\bar p}_{\rm E} = A_1 + 4 \pi^2 A_2 R_{\rm E}^{\rm can},
\label{pE1}
\end{equation}
where $R_{\rm E}^{\rm can}$ denotes the canonically normalized perturbation
series in powers of $g = g(\nu)$:
\begin{eqnarray}
 R_{\rm E}^{\rm can} & = &
\left( \frac{g}{ 2 \pi} \right)^2
{\Bigg \{} 1 + \left( \frac{g}{ 2 \pi} \right)^2
\left[ 2 \beta_0 \ln \left( \frac{{\nu}}{2 \pi T} \right)
+ 6 \frac{A_4}{A_2}
\ln \left( \frac{ \Lambda_{\rm E} }{ \kappa T } \right) \right]
\nonumber\\
&& + \left( \frac{g}{ 2 \pi} \right)^4
{\bigg [} 4 \beta_0^2 \ln^2 \left( \frac{ \nu }{ 2 \pi T}
\right) + 2 \ln \left( \frac{{\nu}}{2 \pi T} \right) 
\left( \beta_1 + 12 \beta_0 \frac{A_4}{A_2}
\ln \left( \frac{ \Lambda_{\rm E} }{ \kappa T } \right) \right)
\nonumber\\
&& + 6 \frac{A_4}{A_2}
{\cal K}_3 \ln \left( \frac{ \Lambda_{\rm E} }{ \kappa T } \right)
+ \delta_{\rm E} 
{\bigg ]} + {\cal O}(g^6) {\Bigg \}} \ .
\label{pE2}
\end{eqnarray}
Here, $\beta_1 = (1/16) (102 - 38 n_f/3)$ is the two-loop
RGE coefficent, and
$\nu = \nu_{\rm E}$ ($\sim 2 \pi T$) is the short-range RScl, and 
\begin{equation}
\kappa = 2 \pi \exp \left[ - \frac{1}{24A_4} (A_3 - 6A_5) \right]
\label{kappa}
\end{equation}
has been introduced for obtaining compact expressions.
The $\mu_f$-dependent function ${\cal K}_3$ has been obtained by 
requiring that 
the dependence of ${\bar p}_{\rm M+G}$ on the factorization scale $\Lambda_{\rm E}$
at order $g^6$ cancel with that in ${\bar p}_{\rm E}$. 
It arises in the following way: inserting expansions (\ref{mE2})-(\ref{l1})
into (\ref{pMGeff}) yields to order $g^6$ the
$\Lambda_{\rm E}$-dependent term
\begin{eqnarray}
&& \frac{1}{T^4} {\bar p}_{\rm M+G} (g^6  \ln \Lambda_{\rm E}) = \nonumber \\
& = & \left( \frac{g}{2 \pi} \right)^6 
\ln \left( \frac{\Lambda_{\rm E}}{2 \pi T} \right) 
24 \pi^2 \left[ - \frac{1}{4} (A_6 + A_4A_7) + 18 (\alpha_{\rm M1} + \alpha_{\rm G}) -
\frac{5}{27} n^2_f {\widetilde \mu}_1^2 \right] \ .
\label{pMGg6LE}
\end{eqnarray}
Cancellation with the corresponding term in ${\bar p}_{\rm E}$ is 
obtained therefore if 
\begin{equation}
{\cal K}_3 = \frac{1}{A_4} \left[ \frac{1}{4} (A_6 + A_4 A_7) 
- 18 (\alpha_{\rm M1} + \alpha_{\rm G}) 
+ \frac{5}{27} n^2_f {\widetilde \mu}_1^2 \right] \ .
\label{K3}
\end{equation}
The remaining (unknown) coefficient at $g^6$
within ${\bar p}_{\rm E}$, which we denote by $\delta_{\rm E}$, is independent of
$\Lambda_{\rm E}$; it may, however, depend on the (small) ratios 
${\overline \mu}_f = \mu_f/(2 \pi T)$.

Since the true values of ${\bar p}_{\rm E}$, $g_{\rm E} ^2$, $ m_{\rm E}^2$
and ${\bar p}_{\rm M+G}$ are RScl-independent, this motivates us
to apply Pad\'e-related resummations separately to the TPS's
of these quantities,
yielding expressions which are much less RScl-dependent
than the corresponding TPS's.
In this context, we note that the RScl independence of
perturbation expansions (with infinitely many terms) 
for the afore-mentioned quantities is only formal, 
because these series diverge.
They diverge strongly at low temperatures when
$g(\nu_{\rm E})$ [and even more so $g(\nu_{\rm M})$] gets large.
Thus, at low temperatures, the formal RScl dependence
of the series does not help in direct evaluations of the TPS's
since the (higher-order-)RScl-dependent 
corrections to the TPS's are large.
Therefore, for this kinematical region, the only way out of this
RScl dependence dilemma seems to be the conversion of 
the perturbative expressions to
other approximants which are much less RScl-dependent 
at each finite order, even at low temperatures. 
And exactly that is the main motivation for
applying Pad\'e(-related) resummations.

Before doing so, we discuss the other uncertainties  of the
TPS's, namely the uncalculated constants 
$\beta_{\rm M2}$ and $\delta_{\rm E}$, and try to estimate their expected size.
The constant $\beta_{\rm M2}$ appears in expression (\ref{pMGeff})
in combination with the term $\ln(\Lambda_{\rm E}/ m_{\rm E})$, where the
latter is expected to dominate. Therefore, $|\beta_{\rm M2}| < 15$ 
represents a rather generous uncertainty  bound for $\beta_{\rm M2}$.
On the other hand, the constant $\delta_{\rm G}$ was estimated in 
Ref.~\cite{DiRenzo:2006nh} to
be $\delta_{\rm G} = -0.2 \pm 0.6$. Here, it will be treated as a
free parameter within the limits $\delta_{\rm G} = 0 \pm 1$.
Thus, we will allow the following variation of the
aforementioned parameters:
\begin{equation}
-1 < \delta_{\rm G} < +1 \ , \qquad
-15 < \beta_{\rm M2} < + 15 \ .
\label{dGdbeM2}
\end{equation}
Concerning $\delta_{\rm E}$ [see Eq.~(\ref{pE2})], 
we note that the parameter 
$\kappa$ was introduced in such a way
that the $\ln[ \nu/(2 \pi T)]$-independent part of
the coefficient at $g^4$ in ${\bar p}_{\rm E}$ is absorbed by a term
proportional to $\ln[\Lambda_{\rm E}/(\kappa T)]$. 
The coefficient at $g^6$ was then organized into
a polynomial in powers of the aforementioned
two logarithms. It is reasonable to expect that
the $\ln[ \nu/(2 \pi T)]$-independent part of
this coefficient is absorbed to a large degree
by a term proportional to $\ln[\Lambda_{\rm E}/(\kappa T)]$.
Therefore, parameter $\delta_{\rm E}$ is expected
to be small and the following variation of this
unknown parameter appears to be rather generous:
\begin{equation}
- | k_2| \ < \ \delta_{\rm E} \ < \
+ | k_2| \ ,
\label{dEest}
\end{equation}
where 
\begin{equation}
k_2 \equiv
6 \frac{A_4}{A_2}
{\cal K}_3 \ln \left( \frac{ \Lambda_{\rm E} }{ \kappa T } \right) \ .
\label{k20}
\end{equation} 
Parameter $\delta_{\rm E}$ depends only on the
small parameters ${\overline \mu}_f$.
On the other hand, the bounds $\pm |k_2|$ will have an
additional slight dependence on temperature $T$ because
we take $\Lambda_{\rm E} = \sqrt{2 \pi T  m_{\rm E}(T)}$ ($\sim g^{1/2} T$).

Formulas in this Section are in close analogy with
those in our previous work \cite{GRp}, involving now
additional (small) parameters ${\overline \mu}_f \equiv \mu_f/(2 \pi T)$.
Furthermore, up to terms $g^4$, they coincide with
those of Ref.~\cite{Braaten:1996ju} when ${\overline \mu}_f=0$.
Further, reexpanding ${\bar p}_{\rm M+G}$ of 
Eq.~(\ref{pMGeff}) in powers of the coupling
parameter $g \equiv g(\nu)$, and adding it to
expansion (\ref{pE1})-(\ref{pE2}) for ${\bar p}_{\rm E}$
while using there the same RScl $\nu$,
gives the same expansion as the one obtained in
Refs.~\cite{Vuo} for $p_{\rm E+M+G}$.

\section{Resummation and Numerical Results}
\label{sec:resum}
Our next step is to apply specific Pad\'e related resummations
to evaluate separately the long-distance (\ref{pMGeff})-(\ref{l1b}) 
and short-distance (\ref{pE1})-(\ref{pE2}) contributions to the pressure.

In principle, we could utilize Pad\'e (P[N/M]) or Pad\'e-Borel
(PB [N/M]) approximants\footnote{For a short description of Pad\'e and
Borel-Pad\'e approximants, see Appendix of Ref.~\cite{GRp}.}
of any possible order [N/M] which is 
compatible with the order $n$ (the highest power of expansion parameter) 
of the TPS: ${\rm N+M} \leq n$.
So we have a certain freedom of choice. We use it for achieving
physically desirable features. These are: 
\begin{itemize}
\item[(a)] Significantly suppressed RScl dependence of both resummed
${\bar p}_{\rm M+G}$ and ${\bar p}_{\rm E}$, with the two RScl's $\nu_{\rm M}$ and
$\nu_{\rm E}$ varying in the regimes $\nu_{\rm M} \sim gT$ and $\nu_{\rm E} \sim 2 \pi T$.
Minimal RScl dependence is achieved in general
for diagonal or near-diagonal approximants (${\rm N} \approx {\rm M}$),
thus we expect that such approximants will be preferred.
\item[(b)] ${\bar p}_{\rm E} + {\bar p}_{\rm M+G}$ should have as little
dependence  on the factorization scale $\Lambda_{\rm E}$ as possible. 
Note that the sum of the original TPS's Eqs.~(\ref{pE1})-(\ref{pE2})
and (\ref{pMGeff}), when expanded in powers of a common $g = g(\nu)$
up to $\sim g^6$,
is completely stable under variation of $\Lambda_{\rm E}$. On the other
hand, the individual parts show significant $\Lambda_{\rm E}$ dependence. 
Since these individual dependences get (individually) changed 
by resummations, we have to 
optimize the approximants in the sense of maximally reducing the artificial 
$\Lambda_{\rm E}$ dependence of ${\bar p}_{\rm E} + {\bar p}_{\rm M+G}$.
\item[(c)] ${\bar p}_{\rm E} + {\bar p}_{\rm M+G}$ should not surpass the value
of $p_{\rm ideal}$ (pressure of the ideal gas), even at low temperatures 
close to the critical temperature $T_c$ (see Ref.~\cite{GRp} for arguments
in this direction).
\end{itemize}
Since ${\bar p}_{\rm M+G}$ is expanded in powers of $(g_{\rm E} ^2/ m_{\rm E})$,
we first have to calculate the EQCD parameters $ m_{\rm E}^2$ and $g_{\rm E} ^2$. We
evaluate them as Pad\'e resummations $P[1/1] (g^2)$ 
of TPS's (\ref{mE2}) and (\ref{gE2}),
thereby banking upon the better convergence behavior of Pad\'e 
approximants and reducing the unphysical RScl dependence dramatically. 

Within our previous paper \cite{GRp}, we have shown for the case
of zero chemical potentials that, among the resummations of the 
Pad\'e (P) and Borel-Pad\'e-type (BP) of the perturbation expansions
(\ref{pMGeff}) for ${\bar p}_{\rm M+G}$ and (\ref{pE2})
for $R^{\rm can}_{\rm E}$, the only physically acceptable ones
in the afore-mentioned sense (a)-(c) are:  

(I) the Pad\'e approximant P[0/3] in
terms of expansion variable $g_{\rm E} ^2/ m_{\rm E}$ for 
${\bar p}_{\rm M+G}/ m_{\rm E}^3$ of Eq.~(\ref{pMGeff}) without the
$\lambda^{(1)}_{\rm E}$-term. Note that, since $g_{\rm E} ^2/ m_{\rm E} \sim g$ and
$ m_{\rm E}^3 \sim g^3$, this emulates partly the diagonal ${\rm P}[3/3](g)$
for ${\bar p}_{\rm M+G}$. Further, the $\lambda^{(1)}_{\rm E}$-term in 
${\bar p}_{\rm M+G}/ m_{\rm E}^3$
of Eq.~(\ref{pMGeff}) is evaluated  according to Eq.~(\ref{l1b}),
making it less RScl-dependent. 

(II) the Borel-Pad\'e approximant BP[1/2] as a function of
$a \equiv g^2/(2 \pi)^2$ for $R_{\rm E}^{\rm can}$ of Eq.~(\ref{pE2}) 
($\Rightarrow {\bar p}_{\rm E}$).

Explicit construction of the Pad\'e's
${\rm P[1/1]}(a)$
for $ m_{\rm E}^2$ and $g_{\rm E} ^2$, ${\rm P[0/3]}(g_{\rm E} ^2/ m_{\rm E})$
for ${\bar p}_{\rm M+G}/ m_{\rm E}^3$, and 
${\rm BP[1/2]}(a)$ for $R_{\rm E}^{\rm can}$ can
be read off from Appendix in Ref.~\cite{GRp}.\footnote{
Fig.~2(b) of Ref.~\cite{GRp}, which shows RScl dependence
of various Pad\'e-related resummations of ${\bar p}_{\rm E}$,
has numerical errors for RScl values
$\nu_{\rm E} \not= 2 \pi T$ due to a mistake in one
of our programs;
corrected curves show that, in addition, P[1/2] resummation
for $R_{\rm E}^{\rm can}$ has an acceptably suppressed RScl dependence.
However, then $p/p_{\rm ideal} > 1$ at
$T \sim 1$ GeV, making it unacceptable 
(where ${\bar p}_{\rm E}$ is from P[1/2] of $R_{\rm E}^{\rm can}$,
and ${\bar p}_{\rm M+G}$ from P[0/3] of ${\bar p}_{\rm M+G}/ m_{\rm E}^3$). 
The conclusion in Ref.~\cite{GRp} that
only BP[1/2] for $R_{\rm E}^{\rm can}$ and P[0/3] for
${\bar p}_{\rm M+G}/ m_{\rm E}^3$ are acceptable remains unaffected.} 
The $\lambda^{(1)}_{\rm E}$-term is not included in the aforementioned
Pad\'e-related resummations, but is evaluated
separately [Eq.~(\ref{l1b})] and added, because it is expected to represent 
diagrams with new, different, topologies.

Since the chemical potentials in expressions (\ref{pMGeff})
and (\ref{pE1})-(\ref{pE2}) are
assumed to be limited in the sense that ${\overline \mu_f} < 1$,
they can be regarded as perturbations to the
${\overline \mu_f}=0$ case \cite{Vuo}. Therefore, we will apply the same 
approximants as in the ${\overline \mu_f}=0$ case of Ref.~\cite{GRp},
i.e., those mentioned above. Furthermore, just as in
Ref.~\cite{GRp}, we will fix the two RScl's $\nu_{\rm E}$ and $\nu_{\rm M}$
according to relations
\begin{equation}
\nu_{\rm E} = 2 \pi T \ ,
\qquad
\nu_{\rm M}^2 =  m_{\rm E}^2(T,{\overline \mu}; \nu_{\rm M}) \
\left[ \equiv  m_{\rm E}^{(0)2}(T,{\overline \mu}) \right] \ ,
\label{RScls}
\end{equation}
where $ m_{\rm E}^2$ is taken to be the Pad\'e approximant ${\rm P[1/1]}(a)$
of expansion (\ref{mE2}), as mentioned before, 
with $g \equiv g (\nu_{\rm M})$. We note that now the
long-distance RScl $\nu_{\rm M}$ will depend on both
the temperature $T$ and the chemical potential $\mu = 2 \pi T {\overline \mu}$.
At sufficiently high temperatures, 
we have $\nu_{\rm M} \sim g T$, cf.~Eq.~(\ref{mE2}).
The factorization scale $\Lambda_{\rm E}$ is chosen to be just in-between
the two RScl's (\ref{RScls}) on the log scale
\begin{equation}
\Lambda_{\rm E} = \sqrt{ \nu_{\rm E} \nu_{\rm M} } =
\sqrt{ 2 \pi T  m_{\rm E}^{(0)} } \ .
\label{LE}
\end{equation}
With all these quantities fixed, there still is a problem of
obtaining results at low temperatures close to the
critical temperature $T_c \approx 0.2$ GeV, where the values of the
long-distance RScl $\nu_{\rm M}$ are much below $1$ GeV. 
At such scales, the usual perturbative ${\overline {\rm MS}}$ couplant
$a(\nu^2) \equiv [g (\nu)/(2 \pi)]^2$ diverges
as a result of the unphysical Landau singularities,
the latter being the consequence of the beta function 
$\beta(a)$ occuring in the form of the (four-loop) truncated perturbation
series (TPS). A partial remedy to the related problem of
unreliability of evolution of $a(\nu^2)$ at low $\nu$
was presented in our previous works \cite{GRp},
where we used Pad\'e ${\rm P[2/3]}(a)$ for $\beta(a)$.
However, the problem of the Landau singularities 
in the low-energy space-like regime persists.
Therefore, in Appendix \ref{app2} we present another
resummation of the ${\overline {\rm MS}}$ four-loop beta function, 
of the Borel-Pad\'e-type (BP).
We show that BP[2/2] and BP[1/3], in ${\overline {\rm MS}}$, result in
evolution which keeps $a(\nu^2)$ finite down to $\nu^2=0$.
The two BP's give mutually similar results. Even more so, when varying
the scheme, e.g., by changing the values of $\beta_2$ and
$\beta_3$ coefficients by about $50\%$, the main
qualitative features of the low-RScl evolution survive. We will
adopt for $\alpha_s (\nu^2, {\overline {\rm MS}}) \equiv \pi a(\nu^2)$
(with the space-like RScl values $q^2 = - \nu^2$)
the reference value 
\begin{equation}
g^2(\nu=m_{\tau})/(4 \pi) \equiv 
\alpha_s (\nu^2=m_{\tau}^2,{\overline {\rm MS}}) = 0.334  \qquad (n_f=2 \ {\rm or} \ 3) \ ,
\label{alsmtau}
\end{equation}
which is approximately the value extracted
from the hadronic $\tau$ decay data 
\cite{Geshkenbein:2001mn,Cvetic:2001sn};
and we will evolve $a(\nu^2)$ by the BP[2/2] beta function (see Appendix \ref{app2},
Figs.~\ref{asbeBPnf3}(a) and \ref{asbeBPnf2}(a) for $n_f=3,2$, 
respectively). Sometimes, for comparison, the evolution by
the BP[1/3] beta function will be used (Figs.~\ref{asbeBPnf3}(b)
and \ref{asbeBPnf2}(b) for $n_f=3,2$).
Details of the definition of BP[i/j] for the beta function
and other details are given in Appendix \ref{app2}.

Most of the following calculations are performed for the case of
two active massless quark flavors ($n_f=2$), in order
to facilitate comparison with the lattice calculations
of Refs.~\cite{Allton:2003vx,Allton:2005gk}. Some of
the calculations will be performed for $n_f=3$ in
order to see the $n_f$ dependence of the results.
We adopt the notations used in Ref.~\cite{Allton:2005gk}: 
\begin{eqnarray}
\Delta p & = & p(T; \mu_f) - p(T; \mu_f=0) \ ,
\label{Dpdef}
\\
\mu_q &=& \frac{1}{2} ({\mu}_u + \mu_d) \ , \qquad
\mu_{I} = \frac{1}{2} ({\mu}_u - \mu_d) \ ,
\label{mus}
\\
n_q &=& \frac{ \partial p}{\partial \mu_q} \ ,
\label{nq}
\\
\chi_q & = & \frac{ \partial^2 p}{\partial (\mu_q)^2} \ , \qquad
 \chi_I =  \frac{ \partial^2 p}{\partial (\mu_I)^2} \ , 
\label{chis}
\end{eqnarray}
where the partial derivatives with respect to $\mu_X$ ($X=q,I$)
are taken at constant $T$ and constant $\mu_Y$ ($Y=I,q$, resp.).
Here, $n_q$ is the quark number density (at $\mu_I=0$);
$\chi_q$ and $\chi_I$ are the quark number and isovector
susceptibilities.
When $n_f=3$, we will take $\mu_s = 0$, as in the lattice
calculations of Refs.~\cite{Fodor:2002km,Csikor:2004ik}.
Further, we will use for the critical temperature
the value $T_c = 0.17$ GeV as in Refs.~\cite{Allton:2003vx,Allton:2005gk},
both for $n_f=2$ and $n_f=3$. 
Unless otherwise stated, the unknown parameters 
$\delta_{\rm G}$, $\beta_{\rm M2}$
[Eq.~(\ref{dGdbeM2})] and $\delta_{\rm E}$ [Eq.~(\ref{dEest})]
will be set equal to their central value zero, and
the RScl's $\nu_{\rm E}$ and $\nu_{\rm M}$ for the short-
and long-distance parts of the pressure will take on the
'canonical' values according to Eq.~(\ref{RScls}).
Numerical calculations were performed using
{\it Mathematica\/} \cite{math}. 

In our calculation of $\Delta p$ we will resum
separately, in the aforementioned way, $p(T,\mu_f)$
and $p(T,\mu_f=0)$, and then subtract the two quantities.
We prefer this approach (instead of trying various
resummations of the perturbation series of the
quantity $\Delta p$) because the directly measured
physical quantities are the full pressures.

\begin{figure}[htb]
\begin{minipage}[b]{.49\linewidth}
 \centering\epsfig{file=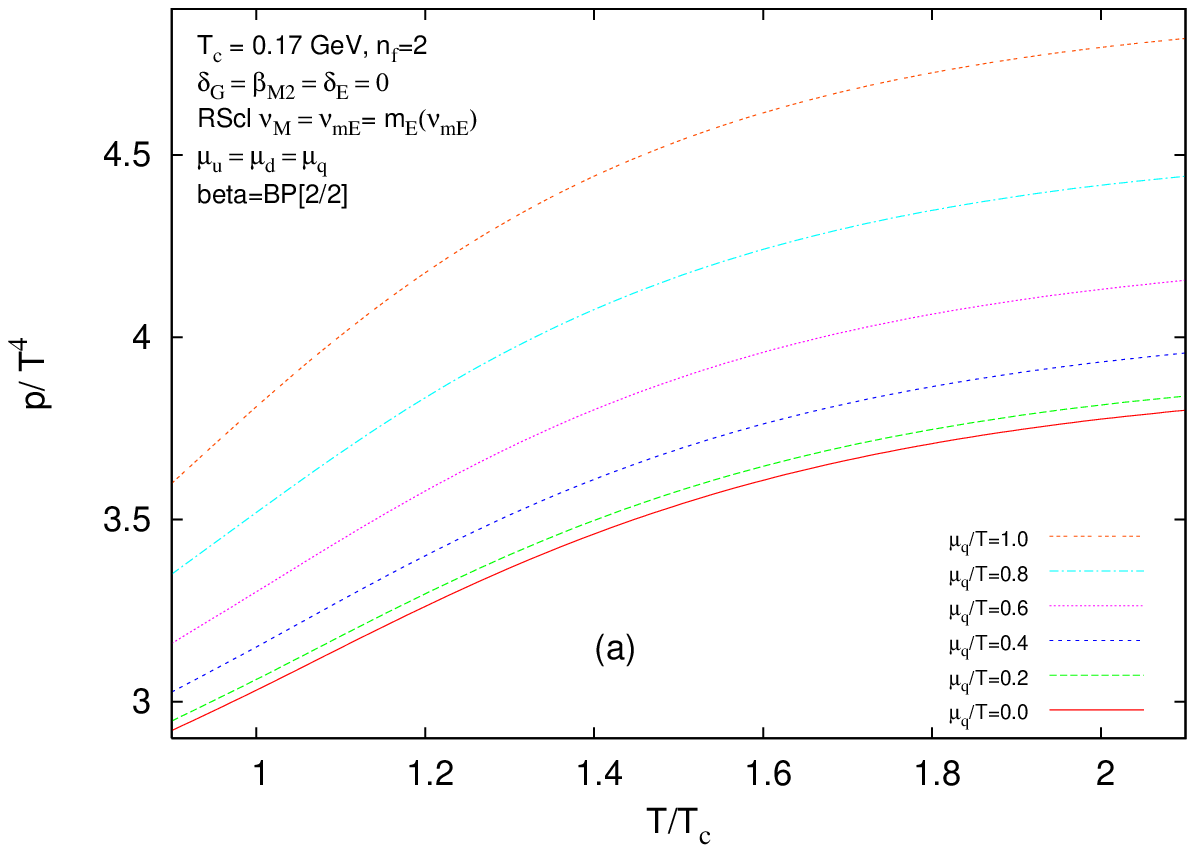,width=\linewidth}
\end{minipage}
\begin{minipage}[b]{.49\linewidth}
 \centering\epsfig{file=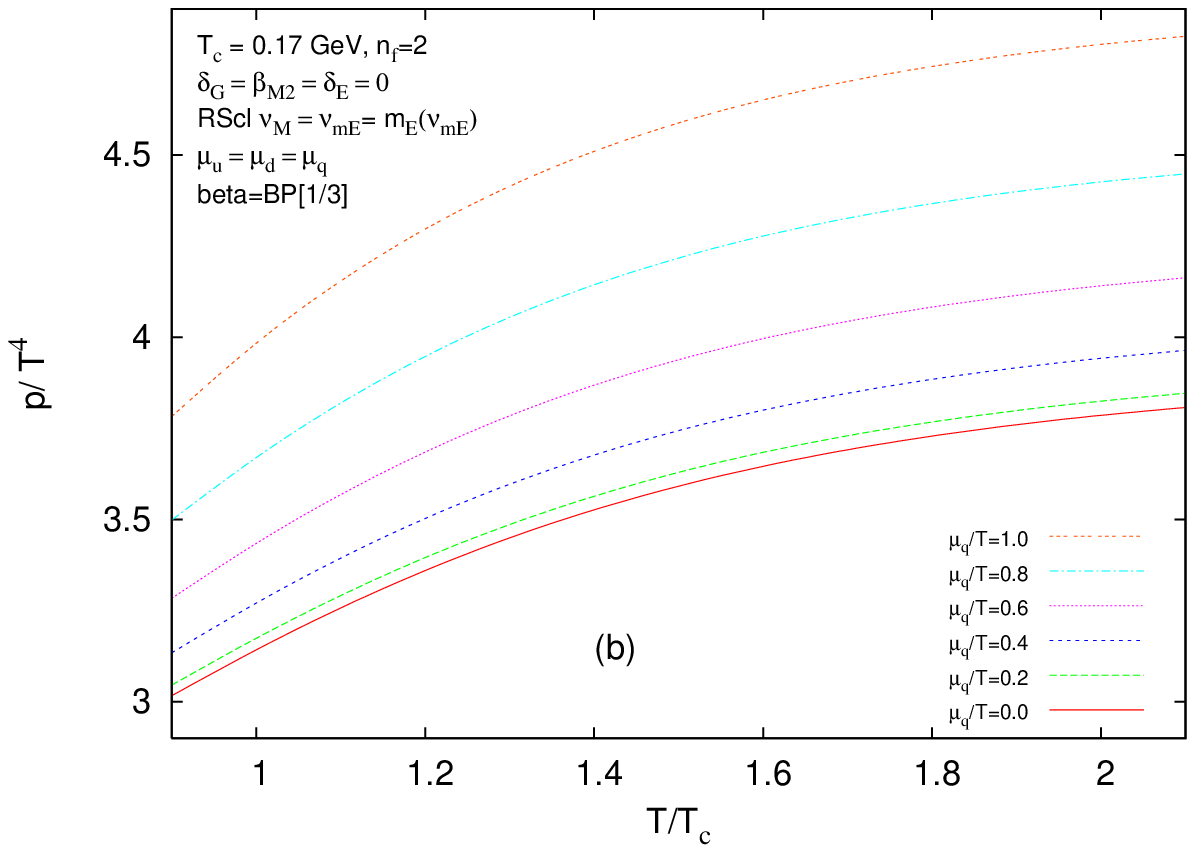,width=\linewidth}
\end{minipage}
\vspace{-0.4cm}
\caption{\footnotesize Pressure $p$ (divided by $T^4$) as a function
of temperature, at various values for the ratios $\mu_q/T$
involving the chemical potential $\mu_q$ when 
(a) Borel-Pad\'e BP[2/2]
and (b) BP[1/3] is used for the beta function $\beta(g_s^2)$.} 
\label{FigpvsT}
\end{figure}
\begin{figure}[htb]
 \centering\epsfig{file=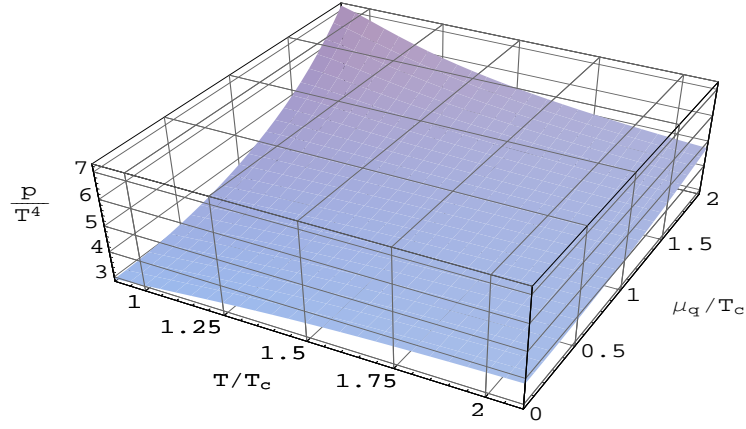,width=10.cm,height=6.5cm}
\vspace{-0.5cm}
\caption{\footnotesize Pressure $p$ (divided by $T^4$) as a function
of $T/T_c$ and $\mu_q/T_c$ (with: $\mu_I=0$). The choice
of parameters, RScl's, and resummation approximants are the
same as in Fig.~\ref{FigpvsT}(a).}
\label{FigpvsTvsmu}
\end{figure}
Numerical resummations, performed in the way described above,
give us for the total pressure $p = {\bar p}_{\rm E} + {\bar p}_{\rm M+G}$
the results presented in Figs.~\ref{FigpvsT}, for various
values of the ratio $\mu_q/T$ (where $\mu_q=\mu_u=\mu_d$),
as a function of temperature in the vicinity of $T_c$.
Comparison of Figs.~\ref{FigpvsT}(a) and (b) further reveals that
the results do not change significantly when the type of the
BP resummation of the beta function is changed. For better visualization,
we present in Fig.~\ref{FigpvsTvsmu} a three-dimensional image,
showing $p/T^4$ as a function of $T/T_c$ and of $\mu_q/T_c$
($\mu_I=0$), for the choice of parameters, RScl's, and resummation
approximants equal to that of Fig.~\ref{FigpvsT}(a). Note, however, that
in Fig.~\ref{FigpvsTvsmu} the second axis is $\mu_q/T_c$
and not $\mu_q/T$ (the latter quantity is kept fixed in
the separate curves of Fig.~\ref{FigpvsT}(a)).

\begin{figure}[htb]
 \centering\epsfig{file=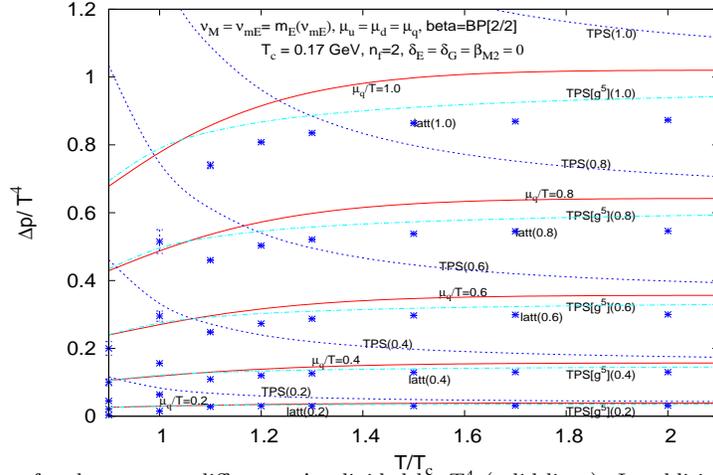,width=10.cm,height=6.5cm}
\vspace{-0.5cm}
\caption{\footnotesize As in Figs.~\ref{FigpvsT},
but now for the pressure difference $\Delta p$
divided by $T^4$ (solid lines). 
In addition, the TPS results are included as dotted lines;
and the lattice calculation
results of Ref.~\cite{Allton:2005gk} are included as crosses,
where the depicted error bars include only specific statistical errors.}
\label{FigDpvsT}
\end{figure}
\begin{figure}[htb]
\begin{minipage}[b]{.49\linewidth}
 \centering\epsfig{file=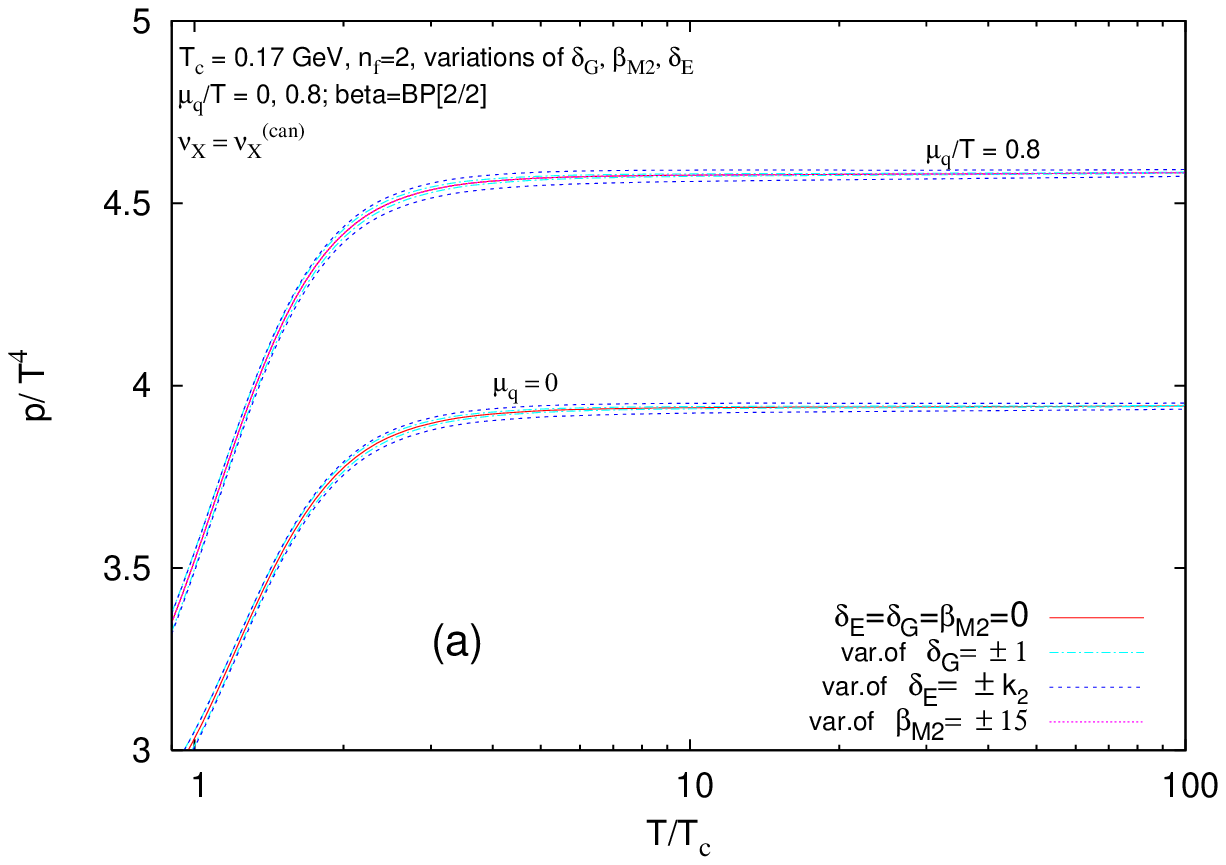,width=\linewidth}
\end{minipage}
\begin{minipage}[b]{.49\linewidth}
 \centering\epsfig{file=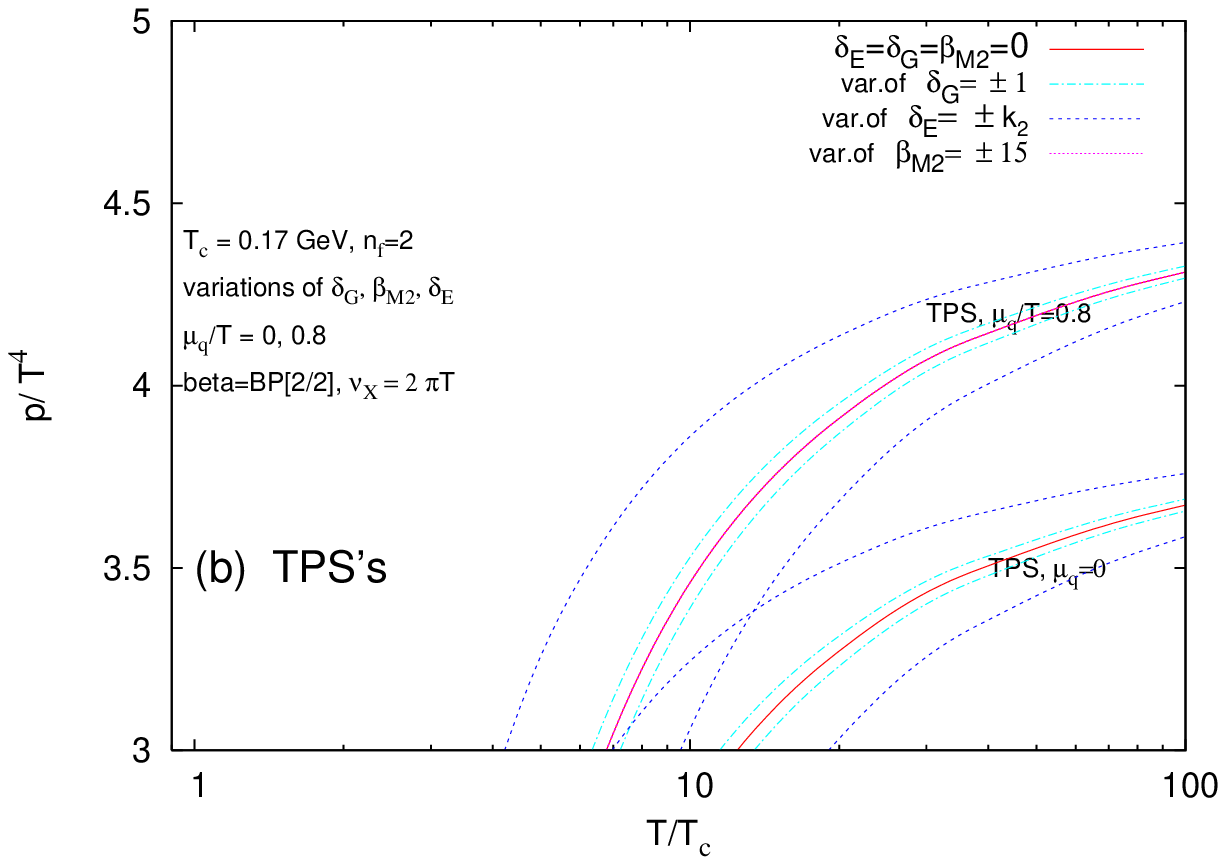,width=\linewidth}
\end{minipage}
\vspace{-0.4cm}
\caption{\footnotesize (a) Pressure $p$ (divided by $T^4$) as a function
of temperature, when the unknown parameters $\delta_{\rm G}$,
$\beta_{\rm M2}$ and $\delta_{\rm E}$ are varied according
to Eqs.~(\ref{dGdbeM2}) and (\ref{dEest}), for two different
values of the ratio $\mu_q/T$: 0.8 and 0; (b) same as in
(a), but for TPS's (with the common RScl $\nu = 2 \pi T$).}
\label{FigpvsTpch}
\end{figure}
\begin{figure}[htb]
\begin{minipage}[b]{.49\linewidth}
 \centering\epsfig{file=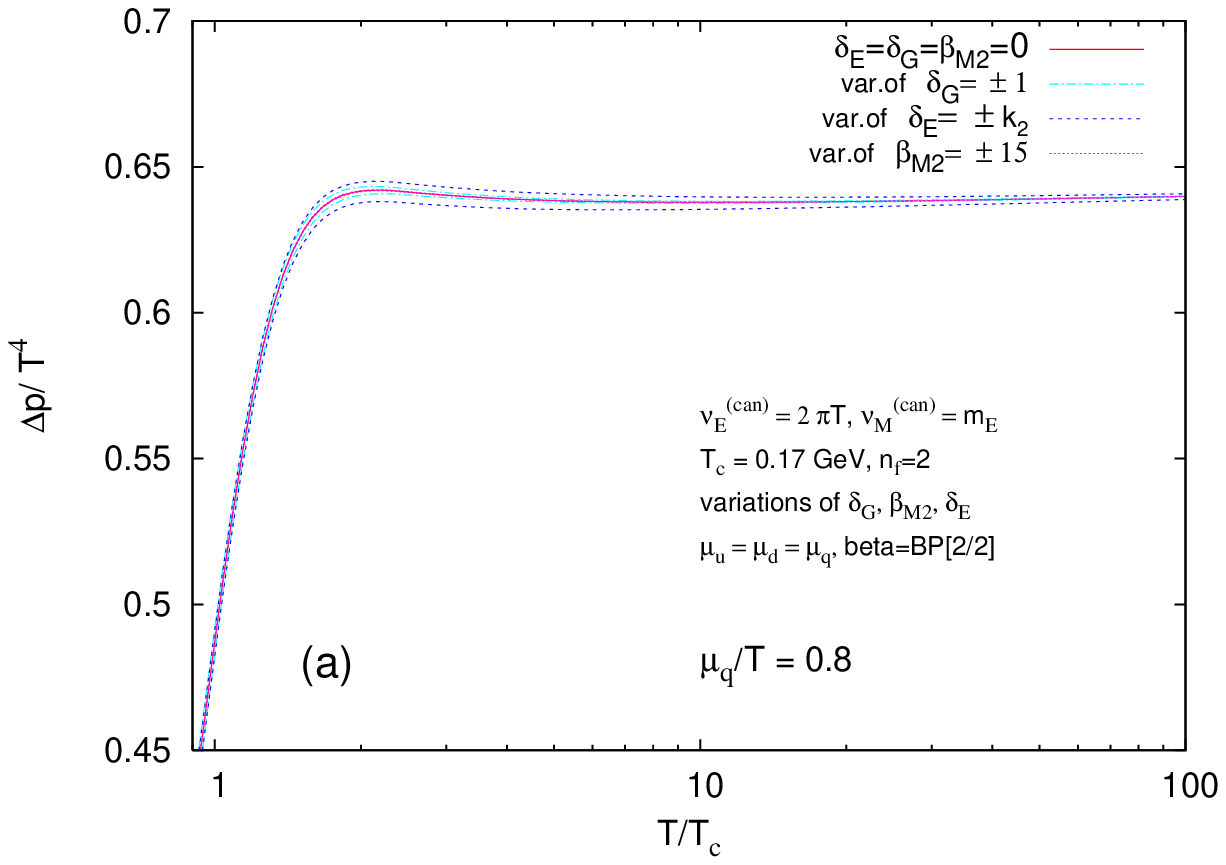,width=\linewidth}
\end{minipage}
\begin{minipage}[b]{.49\linewidth}
 \centering\epsfig{file=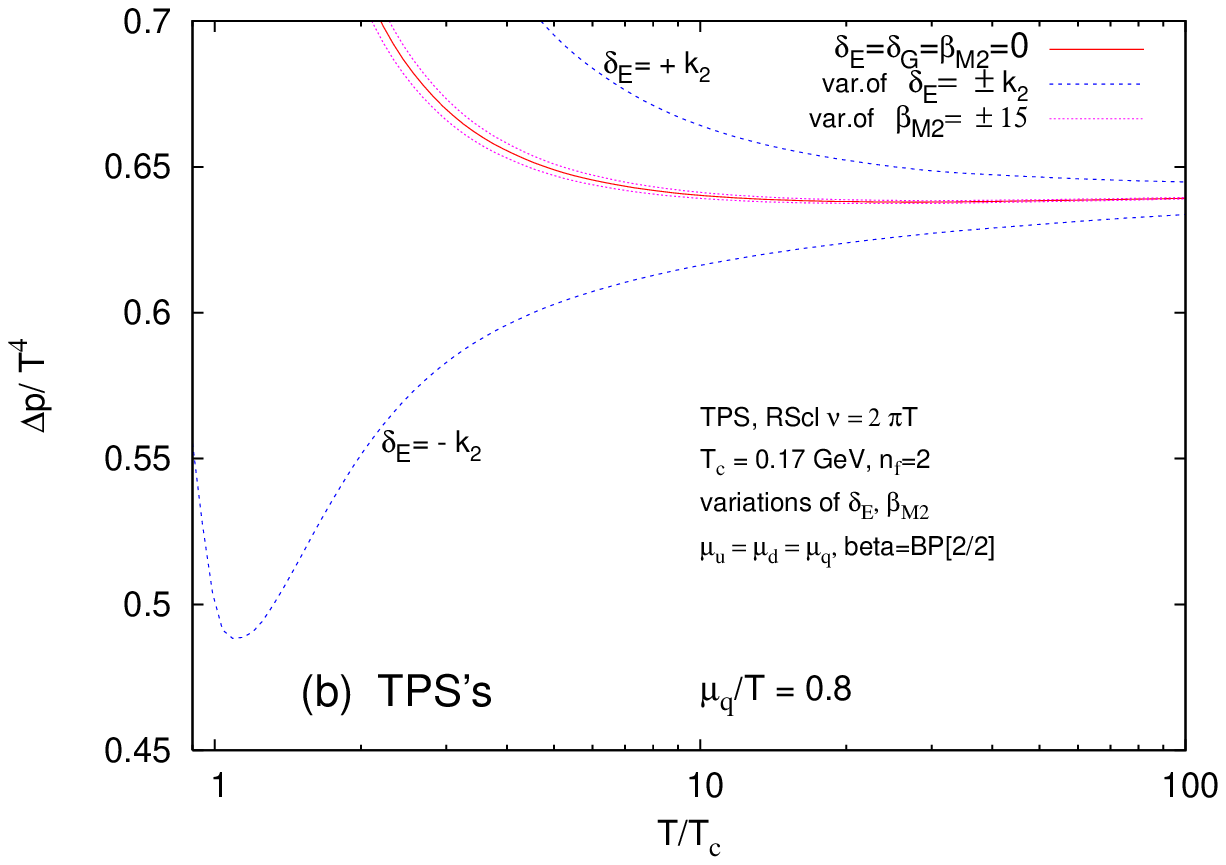,width=\linewidth}
\end{minipage}
\vspace{-0.4cm}
\caption{\footnotesize As in Figs.~\ref{FigpvsTpch},
but now for $\Delta p/T^4$.}
\label{FigDpvsTpch}
\end{figure}

In Fig.~\ref{FigDpvsT} we present the corresponding results for the
pressure difference $\Delta p = p(T; \mu_q) - p(T; \mu_q=0)$,
for five different values of $\mu_q/T$ ($=0.2, 0.4, 0.6, 0.8, 1.$).
We included, for comparison, the results of the
evaluation of the simple truncated perturbation series (TPS) in powers of
$g(\nu=2 \pi T)$ as dotted lines there. These
were obtained by using for ${\bar p}_{\rm E}$ the TPS in powers of
$g^2(\nu=2 \pi T)$, and for ${\bar p}_{\rm M+G}$ the TPS
in powers of $g(\nu=2 \pi T)$. The latter TPS is obtained
by using expansions (\ref{mE2})-(\ref{l1}) in
powers of $g(\nu)$ in expansion (\ref{pMGeff})
for ${\bar p}_{\rm M+G}$ (also in the logarithms there), 
and setting the RScl $\nu= 2 \pi T$.
The unknown parameters $\delta_{\rm E}$, $\delta_{\rm G}$, and 
$\beta_{\rm M2}$, which affect these TPS's at ${\cal O}(g^6)$,
were all set equal to zero here.
In addition, the TPS's truncated at ${\cal O}(g^5)$ are presented,
for the aforementioned five values of $\mu_q/T$.
We note that such types of TPS evaluation (with the
common high RScl $\sim 2 \pi T$) have been often
used in the literature to evaluate $p$ and/or $\Delta p$.
The TPS's presented here do not diverge when approaching even very low
values of temperature because for the
beta functions we use BP[2/2] (or: BP[1/3] in Fig.~\ref{FigpvsT}(b)).
Fig.~\ref{FigDpvsT} shows that our
Pad\'e-related evaluations, while being somewhat higher, 
reproduce the lattice results for $\Delta p/T^4$ to within $20 \%$,
even at low temperatures $T \approx T_c$. 
On the other hand, the TPS results are very unstable under the change of the
truncation order. It appears to be a coincidence that
the TPS's truncated at ${\cal O}(g^5)$ are in good agreement
with the lattice data. Incidentally, the latter TPS's have values similar to
those of ${\cal O}(g^6)$-TPS's with $\delta_{\rm E} = - k_2$ 
[in the latter case, ${\bar p}_{\rm E}$ has the coefficient at
$g^6$ equal zero, cf.~Eq.~(\ref{dEest}) and (\ref{pE2})].
Stated otherwise, the ${\cal O}(g^6)$ TPS's are quite unstable
under the variation of the unknown parameter $\delta_{\rm E}$
while our Pad\'e-related resummations
are quite stable (see also Figs.~\ref{FigpvsTpch} and \ref{FigDpvsTpch}).
Furthermore, the TPS results show a strong RScl dependence, 
as will be shown shortly.

A general remark on the lattice data (included in Fig.~\ref{FigDpvsT}) and their
significance is in order here. The quoted error bars denote only
specific statistical errors and do not represent further uncertainties.   
The aforementioned lattice data have various uncertainties, 
among them the generic uncertainties of lattice calculations coming 
from the continuum limit effects (of up to 10\%, cf.~\cite{Allton:2003vx}),
from the finite size effects (of about 5\%, cf.~\cite{Gliozzi:2007jh}),
and from the uncertainties of the value of $T_c$ (of 2-3\%). 
The results for $\Delta p$ for finite chemical potentials suffer from 
additional problems: since finite
$\mu_q$-values are treated in Ref.~\cite{Allton:2005gk} by applying a 
Taylor expansion in powers of $\mu_q/T$ 
(cf.~Eq.~(3.1) of Ref.~\cite{Allton:2005gk}), one has to calculate the
corresponding coefficients $c_n(T)$ -- more of them when
$\mu_q/T$ is higher. In Ref.~\cite{Allton:2005gk} only the first 
three non-vanishing coefficients ($n=2,4,6$)
have been calculated -- with high instabilities
already for $n=6$ (cf.~Fig.~1 in \cite{Allton:2005gk}). 
The error bars in the lattice data in Fig.~\ref{FigDpvsT}, as well as in
Figs.~\ref{FignqvsT}-\ref{FigchivsT}, present only these specific
uncertainties in calculation of the three $c_n$'s. However,
the terms with $n=8,10,\ldots$ are not included. Consequently, the
lattice results for large $\mu_q/T$ values ($\mu_q/T \geq 0.8$) have to
be considered with reservation. An educated guess leads to the expectation
that the data in Fig.~\ref{FigDpvsT} have an overall uncertainty of 
around 15\%, and probably even higher when $\mu_q/T \approx 1$. 
Therefore, it is fair to say that our predictions are in reasonable 
agreement with lattice data down to $T = T_c$, at least as long as 
$\mu_q < T$. Analogous statements
are valid for the comparison with lattice data in Figs.~\ref{FignqvsT} and 
\ref{FigchivsT} (see later).
What we do not yet understand is the apparent systematics of the lattice
results: they lie systematically below our predictions, in particular
for high $\mu_q$ values. Whether this demonstrates a lattice artefact,
possibly connected with the rather large bare quark mass used there, has to be
further investigated. In this context, we further note that the difference between
our and lattice results could not be significantly reduced
by choosing different values for the unknown parameters $\delta_{\rm E}$,
$\beta_{\rm M2}$, $\delta_{\rm G}$
(which were set equal to zero in Fig.~\ref{FigDpvsT}), 
at least if varying them within the generous ranges specified in 
Eqs.~(\ref{dGdbeM2}) and (\ref{dEest}).
In fact, variation of these parameters can decrease our results at 
$1<T/T_c<2$ by less than 1\% (see Fig.~\ref{FigDpvsTpch} later).

In Fig.~\ref{FigpvsTpch}(a) we present variations of
our results for $p/T^4$ in a wide temperature regime when the
unknown parameters $\delta_{\rm G}$,
$\beta_{\rm M2}$ and $\delta_{\rm E}$ are varied according
to Eqs.~(\ref{dGdbeM2}) and (\ref{dEest}), for two
different fixed values of the ratio $\mu_q/T$ ($=0.8, 0.$).
In Fig.~\ref{FigpvsTpch}(b), the analogous results for the aforementioned
simple TPS's are shown. We see that our results are
remarkably stable under the rather generous variations of the
three unknown parameters, whereas this is definitely not the case
with the TPS's. 
The dependence on the unknown parameters
$\delta_{\rm E}$ and $\delta_{\rm G}$ is strong in the TPS's, 
while the Pad\'e-related resummation results are almost
independent of them.
The dependence on the parameter $\beta_{\rm M2}$ is too weak to
be seen.

In Figs.~\ref{FigDpvsTpch}(a),(b) we present $\Delta p/T^4$
in the way completely analogous to the presentation of $p/T^4$
in Figs.~\ref{FigpvsTpch}(a),(b). The conclusions for
the (in)stability of the calculated $\Delta p$ under the variation
of the unknown parameters are similar to those for $p$.
The independence of the parameter
$\delta_{\rm G}$ in the TPS's in Fig.~\ref{FigpvsTpch}(b) is a 
direct consequence of the $\mu_q$ independence of $\delta_{\rm G}$.
The dependence on the parameter $\beta_{\rm M2}$ is weak
in the TPS's, and too small to be seen in the Pad\'e-type resummation.
However, while there is almost no dependence on the unknown
parameter $\delta_{\rm E}$ in the Pad\'e-related resummation results, 
the dependence on $\delta_{\rm E}$ is quite drastic in the TPS's.
This behavior also makes plausible the fact that, by
adjusting the value of the unknown parameter $\delta_{\rm E}$,
we can, in a way, fine-tune the TPS results to come close to the
lattice results (see also Fig.~\ref{FigDpvsT}).

\begin{figure}[htb]
\begin{minipage}[b]{.49\linewidth}
 \centering\epsfig{file=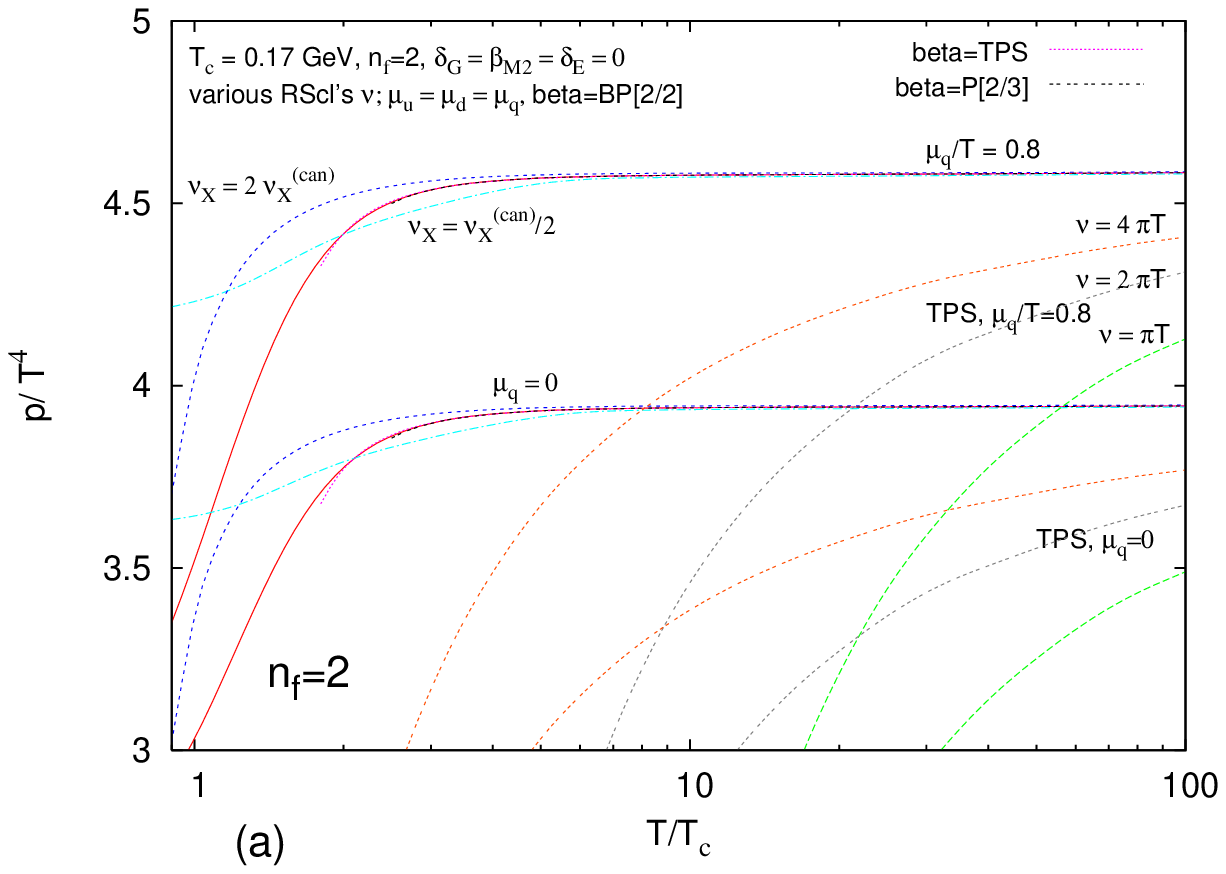,width=\linewidth}
\end{minipage}
\begin{minipage}[b]{.49\linewidth}
 \centering\epsfig{file=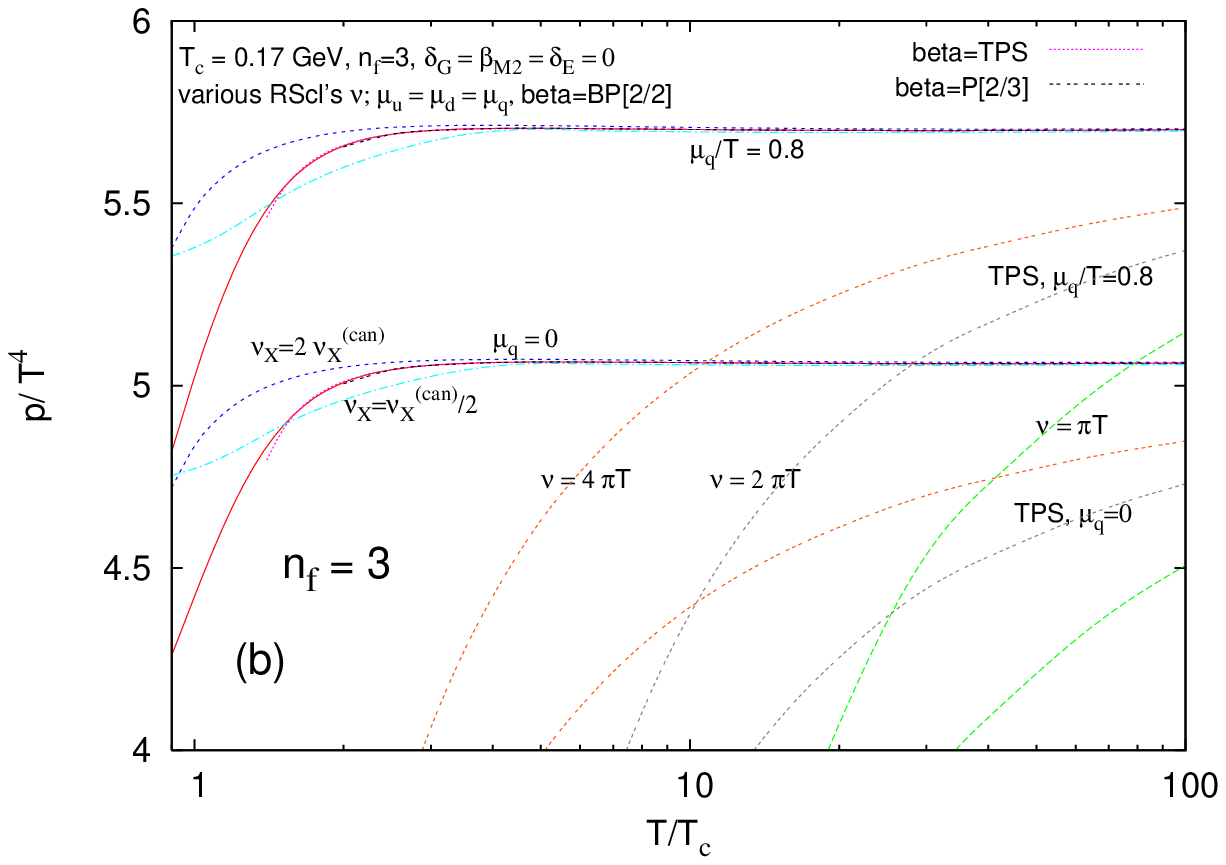,width=\linewidth}
\end{minipage}
\vspace{-0.4cm}
\caption{\footnotesize Pressure $p/T^4$ as a function
of temperature, when the renormalization scales $\nu_{\rm E}$
and $\nu_{\rm M}$ are varied by factor two
around the canonical values (\ref{RScls}): (a) when $n_f=2$; 
(b) when $n_f=3$. The values of the ratios $\mu_q/T$ are
either $0.8$ or zero. Included are also the corresponding
sets of curves with the TPS evaluation.} 
\label{FigpvsTRScl}
\end{figure}

\begin{figure}[htb]
\begin{minipage}[b]{.49\linewidth}
 \centering\epsfig{file=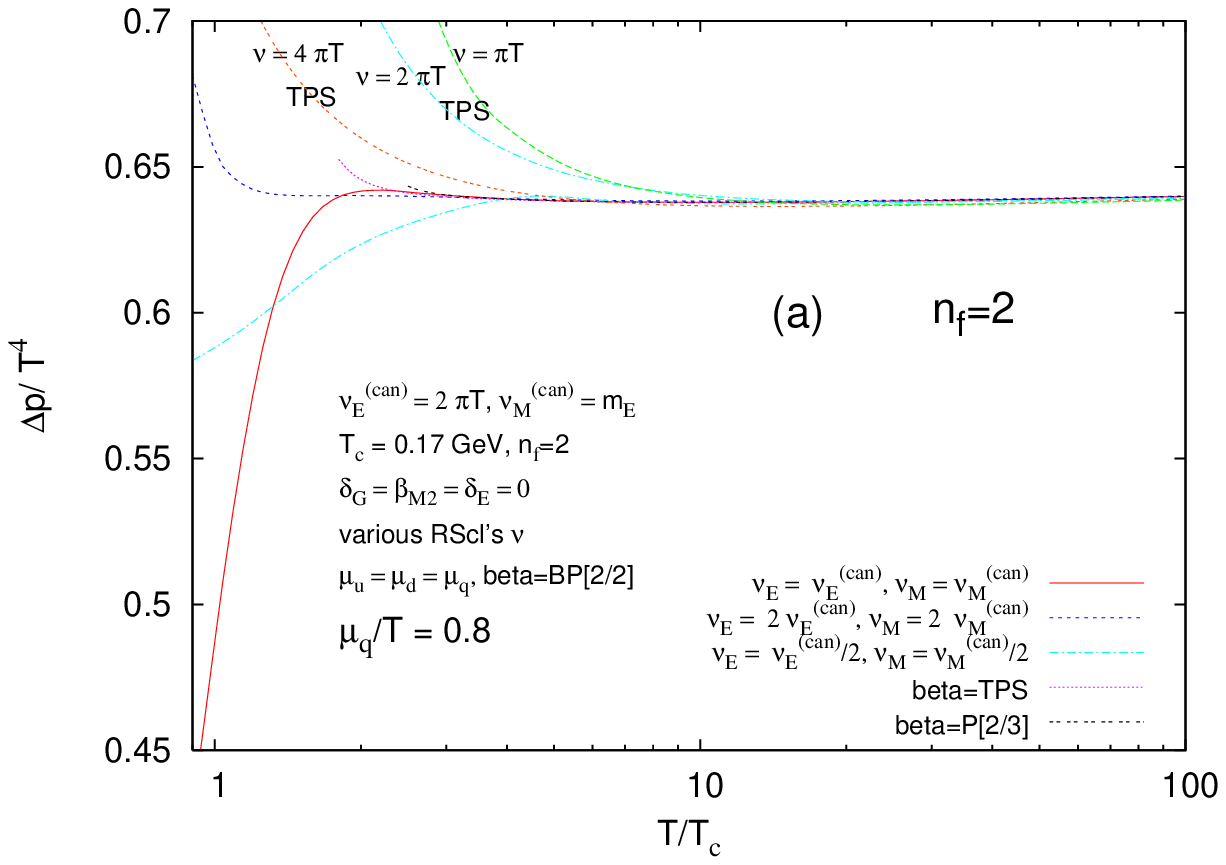,width=\linewidth}
\end{minipage}
\begin{minipage}[b]{.49\linewidth}
 \centering\epsfig{file=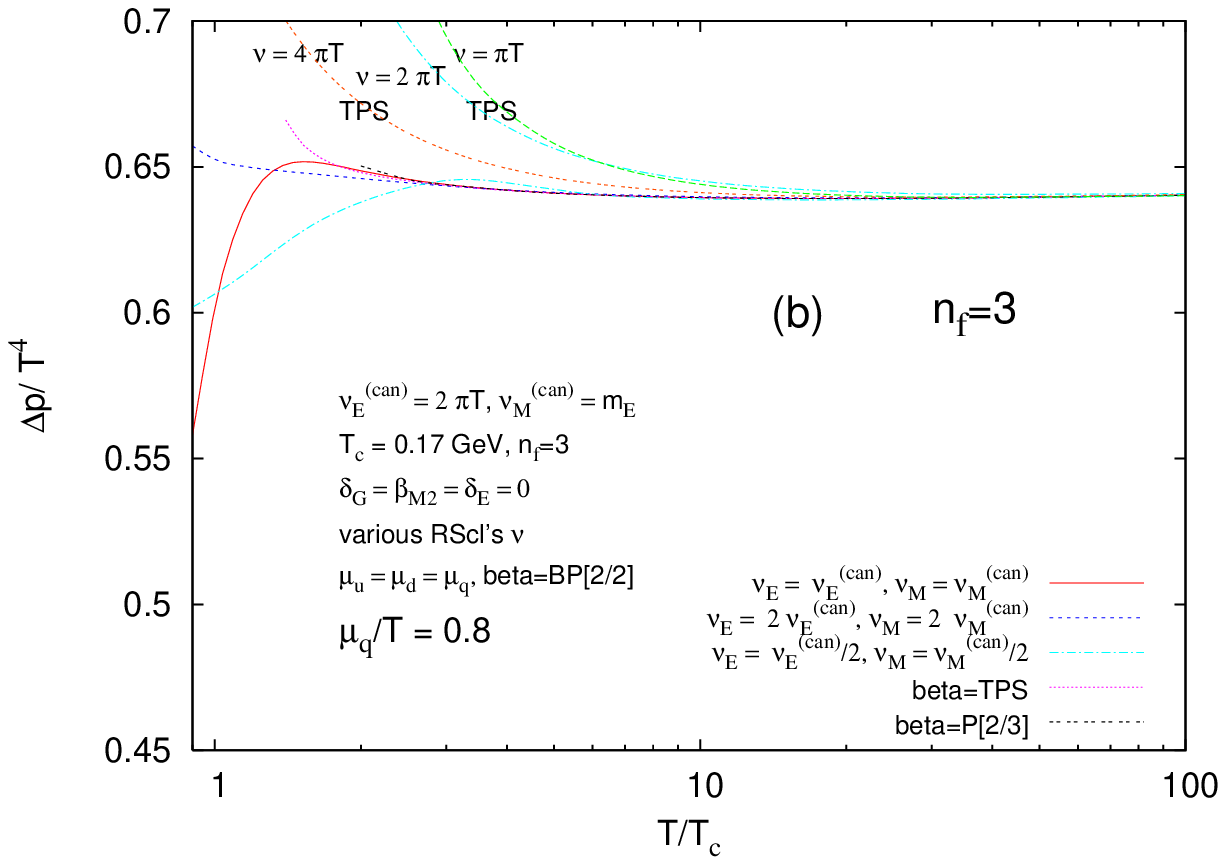,width=\linewidth}
\end{minipage}
\vspace{-0.4cm}
\caption{\footnotesize As Figs.~\ref{FigpvsTRScl}, but for
$\Delta p/T^4$ instead of $p/T^4$. The chemical potential
$\mu_q$ has the values $\mu_q = 0.8 T$.}
\label{FigDpvsTRScl}
\end{figure}

\begin{figure}[htb]
 \centering\epsfig{file=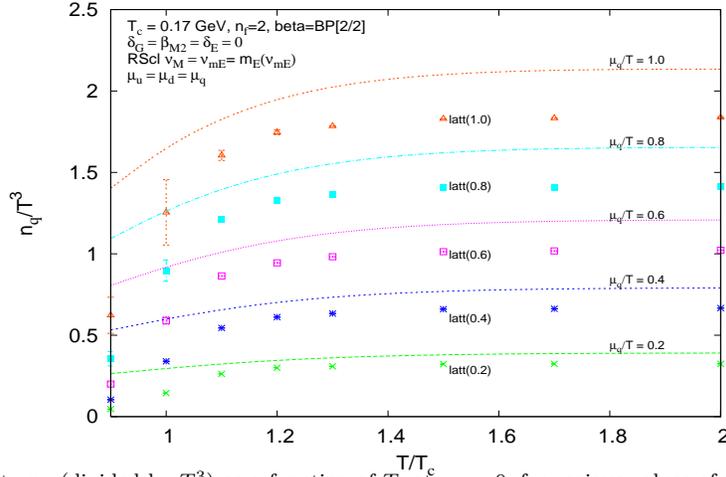,width=10.cm,height=6.5cm}
\vspace{-0.5cm}
\caption{\footnotesize Quark number density $n_q$ (divided by
$T^3$) as a function of $T$, at $\mu_I=0$, for various
values of the ratio $\mu_q/T$. Our results
are in the form of curves. Included as points are
the corresponding results of the lattice calculation
of Ref.~\cite{Allton:2005gk}.}
\label{FignqvsT}
\end{figure}

\begin{figure}[htb]
\begin{minipage}[b]{.49\linewidth}
 \centering\epsfig{file=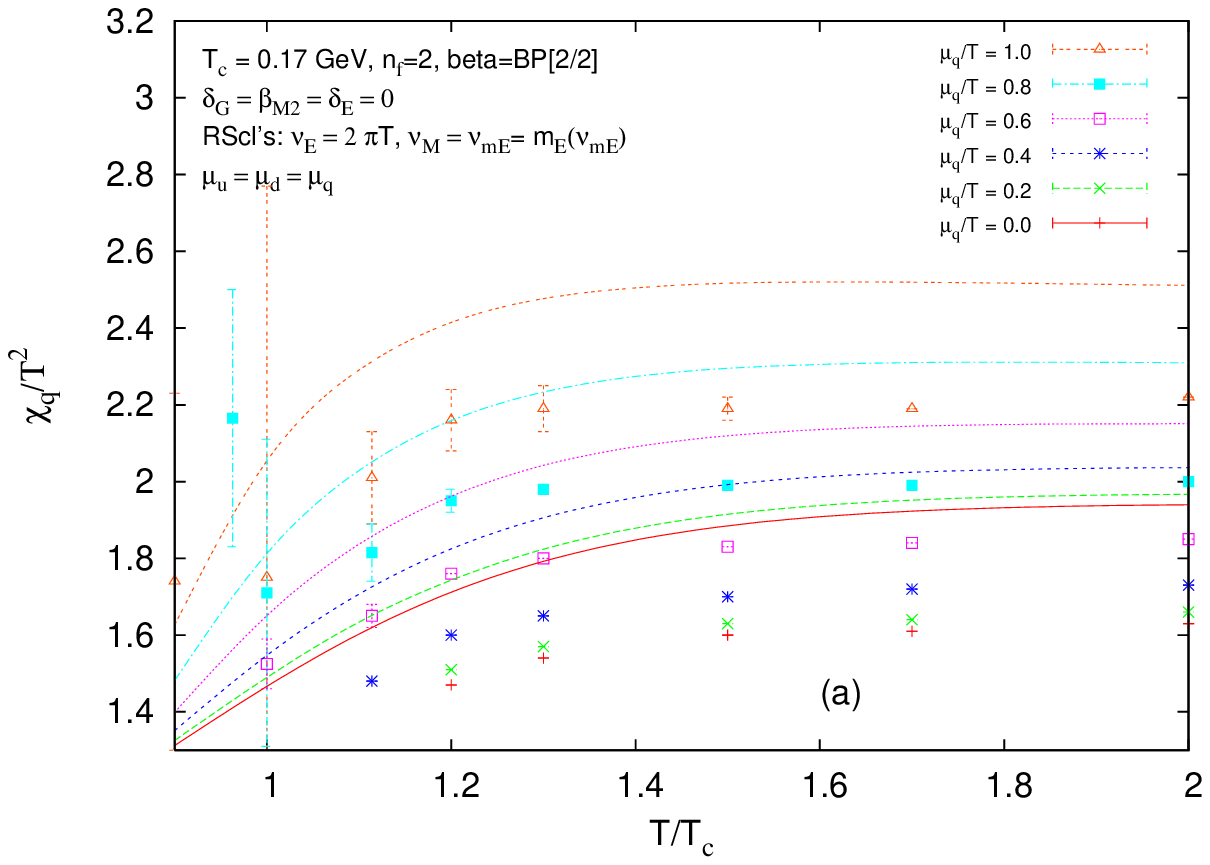,width=\linewidth}
\end{minipage}
\begin{minipage}[b]{.49\linewidth}
 \centering\epsfig{file=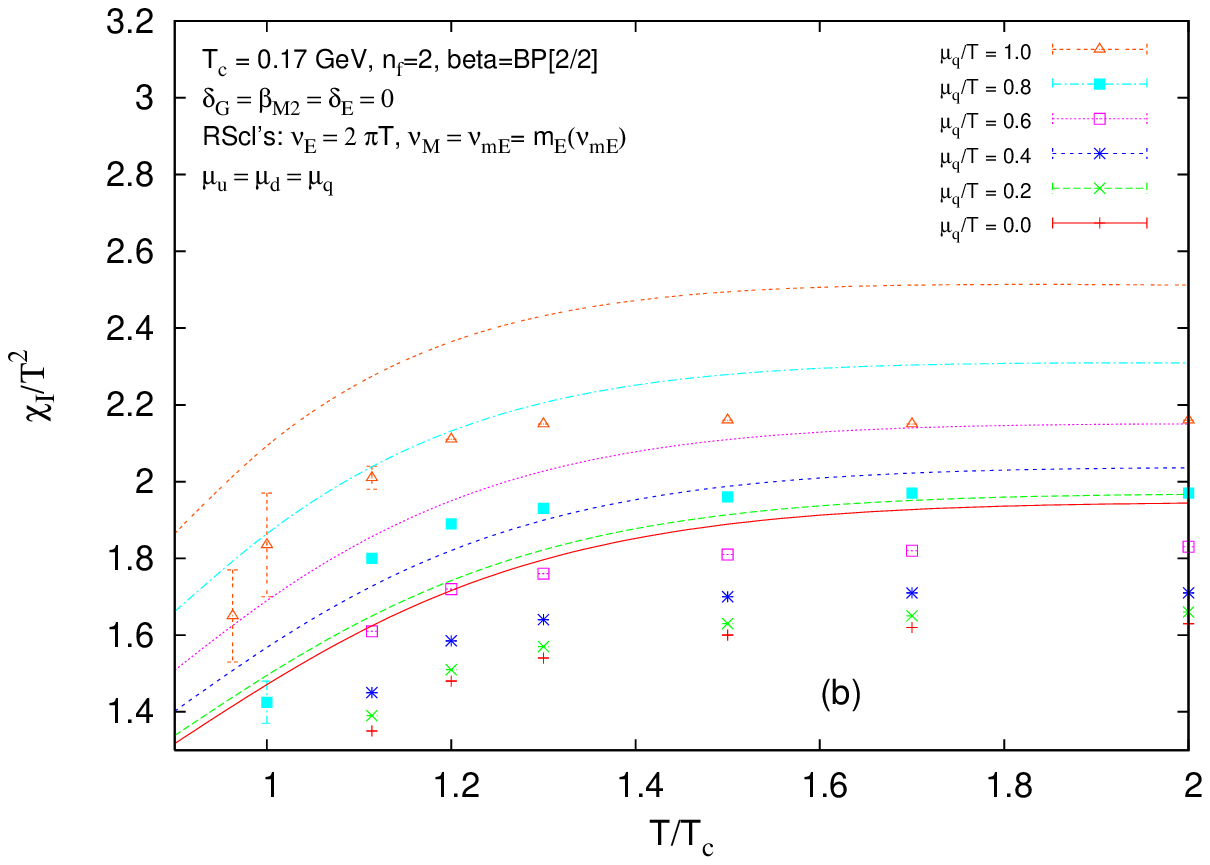,width=\linewidth}
\end{minipage}
\vspace{-0.4cm}
\caption{\footnotesize The susceptibilities (a) $\chi_q$ and
(b) $\chi_I$ as a function of temperature, for various
values of the ratios $\mu_q/T$ ($\mu_I=0$). Our results
are in the form of curves. Included are
the corresponding results 
of the lattice calculation of Ref.~\cite{Allton:2005gk},
in the form of points (some with error bars).}
\label{FigchivsT}
\end{figure}

After having shown that our results are fairly insensitive to
the still existing unknown parts of the perturbation series (at $\sim g^6$), 
we now come to the most important results of our
approach: the stability under variation of the (two) RScl's, even at
very low temperatures. This is manifested in 
Figs.~\ref{FigpvsTRScl} and \ref{FigDpvsTRScl}.

In Figs.~\ref{FigpvsTRScl}(a),(b), we present the behavior of our evaluated 
results for $p/T^4$ when the RScl's $\nu_{\rm E}$ and $\nu_{\rm M}$ 
are varied by factor two
around the canonical values (\ref{RScls}), for
$n_f=2,3$, respectively.
Two sets of curves are given, for $\mu_q/T=0.8$ and zero,
respectively. In addition, the corresponding sets of curves
for the aforementioned simple TPS's are shown, where
now the common RScl $\nu$ is varied from $\pi T$ to $4 \pi T$.
We see that our evaluated results for $p$ are
much more stable under the variation of RScl than
the TPS results, down to very low temperatures
$T \approx T_c$. This is the same conclusion
as the one obtained in our previous work
\cite{GRp} for the case of zero chemical potential ($\mu_q=0$).
For additional comparisons, 
we included in Figs.~\ref{FigpvsTRScl} the
results of our Pad\'e-related evaluation of $p/T^4$ with canonical
RScl's (\ref{RScls}) when the ${\overline {\rm MS}}$ beta function $\beta(a)$
is (four-loop) TPS, and when it is Pad\'e  ${\rm P[2/3]}(a)$.
We see that the results in such
cases, when they exist, almost coincide with the
solid-line curves, i.e., with those with $\beta(a) = {\rm BP[2/2]}(a)$.
However, due to the Landau singularities of $a(\nu^2)$ at low
RScl's in the aforementioned cases of $\beta =$ TPS
or P[2/3] (cf.~Appendix \ref{app2}), the corresponding curves 
exist (i.e., do not blow up) only down to 
$T_{\rm min} \approx 1.8 T_c, 2.5 T_c$, respectively
(when $n_f=3$: $T_{\rm min} \approx 1.4 T_c, 2.0 T_c$, respectively).
We note that the curve with $\beta=$ P[2/3] and $\mu_q=0$
in Fig.~\ref{FigpvsTRScl}(b) corresponds to the
central curve with $n_f=3$ in Fig.~18 of our previous
work \cite{GRp} and to the upper solid-line curve
of Fig.~19 of that work.\footnote{
We mention that a numerical mistake was committed in
the mentioned curve of Ref.~\cite{GRp}, in that the
power of $(g_{\rm E} ^2/ m_{\rm E})$ in the program there was taken to
be five instead of three [cf.~Eq.~(\ref{l1b})]. The
curve with $\beta=$ P[2/3] and $\mu_q=0$ in the
present Fig.~\ref{FigpvsTRScl}(b) now represents
the corrected version of the mentioned curve.}

Figs.~\ref{FigDpvsTRScl}(a),(b) contain similar results for $\Delta p/T^4$
(here only for the value $\mu_q/T=0.8$).
The conclusions about the RScl dependence of the results
for $\Delta p/T^4$ are virtually the same as for $p/T^4$.

It is exactly this independence of $\nu$ at $T$ down to about $2 T_c$ which
makes our comparison with lattice data (cf.~Fig.~\ref{FigDpvsT}) 
much more trustworthy
than the simple TPS evaluation.

In the remaining part we present results for derived quantities,
specifically quark number densities and susceptibilities.

Fig.~\ref{FignqvsT} contains results for the quark
number density (\ref{nq}) for various values of
$\mu_q/T$ ($\mu_I=0$). Included are the corresponding results 
of the lattice calculation of Ref.~\cite{Allton:2005gk},
in the form of points (some with error bars).
The values were obtained by numerical differentiation
of our results for $p$ with respect to $\mu_q$ (with
$\mu_I=0$ and $T$ constant). Again, we see that our
results in general agree with the lattice results 
to within $20\%$, even at low temperatures $T \approx T_c$. 

Finally, in Figs.~\ref{FigchivsT}(a),(b) we present
the values for the susceptibilities $\chi_q$ and
$\chi_I$ [cf.~Eq.~(\ref{chis})] as a function of
temperature, for various values of the ratios
$\mu_q/T$, while keeping $\mu_I=0$. The results
are presented as various curves, and were obtained,
at a given $T$, by numerical
evaluation of the double derivatives of our results
for the pressure $p$
with respect to $\mu_q$ (at constant $\mu_I=0$)
and with respect to $\mu_I$ (around $\mu_I=0$, at
constant $\mu_q$) \cite{math}.
In Figs.~\ref{FigchivsT}(a),(b) we included the
corresponding results 
of the lattice calculation of Ref.~\cite{Allton:2005gk},
as points with error bars. Our curves, in general, give
results which are by roughly $20\%$ higher than the lattice results.

In addition to the aforementioned susceptibilities,
the mixed susceptibility $\chi_{ud}$ which
is related to the previous two by
\begin{equation}
\frac{ \chi_I - \chi_q }{T^2} = - 4 \frac{\chi_{ud}}{T^2}
\label{chiud}
\end{equation}
has been a subject of interest in the literature.
Our numerical results give for the above quantity (\ref{chiud}),
at $\mu_I=0$ (and $n_f=2$), values of about $4 \times 10^{-3}$ at 
$\mu_q=0$ and at temperatures $1 < T/T_c < 2$
[these values turn negative ($\sim - 10^{-3}$) when $\mu_q/T=0.2$].
The authors of Ref.~\cite{Allton:2003vx} obtained,
by their lattice calculations,
for the above quantity (\ref{chiud}) at $\mu_I=\mu_q=0$ (and $n_f=2$)
decreasing values as the temperature
increases from $T_c$ to $1.5 T_c$, and at $T/T_c=1.36$ they
found a value of $6.6 \times 10^{-3}$.
Our results for this quantity are roughly in agreement with
the lattice results of Refs.~\cite{Allton:2003vx,Bernard:2002yd},
and with the hard thermal loops (HTL) perturbative estimates of 
Ref.~\cite{3lbebMS}. The latter estimates give for the quantity
(\ref{chiud}), at $1 < T/T_c < 2$, values 
between $5 \times 10^{-3}$ and $1.4 \times 10^{-2}$, 
when using there our values of $a(2 \pi T)$ 
(with the beta function being BP[2/2]).
The lattice quenched results of Refs.~\cite{GG} give for this 
quantity values $\sim 10^{-6}$, i.e., three orders of magnitude lower. 

Further, we performed numerical calculations of
$p$, $\Delta p$ and $n_q$ in the case of three active
flavors $n_f=3$, with $\mu_u = \mu_d = \mu_q$ ($T$-independent)
and $\mu_s=0$, with our approach described above.
Comparisons with the corresponding lattice calculations
of Ref.~\cite{Fodor:2002km} (their Figs.~3 and 6) revealed that,
at their values of $\mu_{\rm B} \leq 0.53$ GeV
($\mu_{\rm B} \equiv 3 \mu_q, n_{\rm B} \equiv n_q/3$), 
our results for $\Delta p$ and $n_{\rm B}$ are
somewhat higher than theirs, by less than $20\%$
at $T/T_c \geq 1.5$, and by $30\% - 40\%$ at $T/T_c = 1.1$.
Direct comparison with the lattice results of Ref.~\cite{Csikor:2004ik}
is not possible, as no continuum limit correction factor ($c_{\mu}$)
was applied there. In Ref.~\cite{Fodor:2002km}, an estimated
correction factor $c_{\mu} = 0.446$ was applied to $\Delta p$.

\section{Summary}
\label{sec:Summary}
Within the present paper we extended our recent approach \cite{GRp}
for improving perturbative expressions for the quark-gluon pressure
(obtained by FTPT) to the case of
finite (but still small) quark densities. Thereby, the main aim was to find
a consistent method for extrapolating the FTPT-based results down to
temperatures as low as $T_c$ ($\approx 200$ MeV). For such low energies, the 
original TPS's are plagued by huge uncertainties, stemming mainly from
their strong RScl dependence  which itself is partially connected with the 
occurrence of, at least, two different energy scales contributing to the
thermodynamic potential under investigation. Therefore, 
simple FTP series do not permit
a reliable comparison with existing (low energy) lattice data. Such
a check can only be performed if the wild RScl dependence is sufficiently
tamed. Our method allows such a taming. It rests mainly on two crucial
points:  Firstly, we  performed a careful separation of the low-energy from
the high-energy contributions to the pressure, which are responsible for
the (in principle measurable) long- and the short-range behavior, respectively.
In this way, we can clarify which values of the RScl are the natural
ones -- they are different for the two parts. Secondly, for each
of these contributions we identified Pad\'e-related
approximants which -- besides showing
other physically desirable features -- led to (almost) RScl-stable
expressions and thus to predictions which can be safely
used down to low temperatures. However,
the use of the approximants for the low-energy (long-range) contributions
at very low temperatures $\sim T_c$
is only possible if the unphysical
perturbative Landau singularities of the QCD coupling parameter
at low energies are eliminated; we did this by using similar
Pad\'e-related approximants for the renormalization-group beta function.
As a result, we demonstrated that the obtained expressions for the
pressure $p$ and the difference $\Delta p = p(\nu_f) - p(0)$
are fairly insensitive to the (as yet) unknown part of 
contributions of ${\cal O}(g^6)$ and to variations of the RScl's,
both of these features being in stark contrast with the TPS expressions. 

Our expressions show a surprisingly good agreement with lattice data -- 
not only for the pressure and its 
$\mu_f$ dependence but also for derived quantities, in particular 
susceptibilities. In this context, we note that the unknown 
relative deviations of the low temperature lattice results for $p$ and $\Delta p$ 
from the true values are expected to be roughly in the range of 10-20\%.
This is due to the well-known lattice artefacts, in particular the ones 
connected with finite-$\mu_f$ effects (truncated Taylor expansion in $\mu_f/T$).
Our Pad\'e-related evaluations give for $\Delta p/T^4$ results which are 
by not more than 20\% higher than the lattice results when 
$\mu_q/T \leq 0.8$ and $1 < T/T_c < 2$, see Fig.~\ref{FigDpvsT}.

Our approach is valid only for values 
of the chemical potentials smaller than the
temperature, because only in this case dimensional reduction can be 
applied. Fortunately, present day heavy ion collisions are probing 
the region with values $\mu_f \alt 50$ MeV, which are small compared to the
temperatures $T \alt 5 T_c$ typically involved. For other kinematic
situations, in particular for small $T$ and larger chemical potentials,
different reorganizations of perturbation expansions
are necessary and have been applied in the literature, the most prominent
one being the hard dense loop approximation which is genuinely
four-dimensional but based on a non-local effective action \cite{Taylor}.
Recently, a purely diagrammatic calculation of the perturbative QCD
pressure (i.e., without involving any effective theory) has been
performed \cite{Ipp} which, at least in principle, should be valid for
all kinematic regions. As it should be expected, these results -- when
applied to high temperatures  and (relatively) low chemical potentials --
are in accordance with those of the dimensional reduction approach.

\begin{acknowledgments}
We are thankful to A.~Vuorinen and E.~Laermann for several helpful comments.
This work was supported in part by Fondecyt (Chile) grant No.~1050512 (G.C.)
and in part by Fondecyt (Chile) International Cooperation
grant No.~7050233 (R.K.). G.C. would like to thank the Department
of Physics, Bielefeld University, for the kind hospitality offered to him
in July 2006.
\end{acknowledgments}

\appendix

\section{Relevant coefficients for the pressure at finite
chemical potential}
\label{app1}
Here we compile expressions for parameters
$A_j$, $ m_{\rm E}$ and $g_{\rm E}$ which we obtained
from expressions of Ref.~\cite{Vuo}
by application of the method of separation
of the long-distance from short-distance
contributions (i.e., introduction of factorization
scale $\Lambda_{\rm E}$: $ m_{\rm E} < \Lambda_{\rm E} < 2 \pi T$), as explained in
the beginning of Sec.~\ref{sec:PEs}.
We denote by $\nu$ the renormalization scale (RScl), 
and $g \equiv g(\nu)$ in ${\overline {\rm MS}}$ scheme.
Other notations used in this appendix are:
\begin{eqnarray}
&& {\overline \mu}_f = \frac{\mu_f}{2 \pi T} \ ,
\qquad z_f = \frac{1}{2} - {\rm i} {\overline \mu}_f \ ,
\label{not1}
\\
&& {\widetilde \mu}_k  =  \frac{1}{n_f} \sum_{f} {\overline \mu}_f^k ,
\ 
{\widetilde \aleph}^{(k)}(z) = \frac{1}{n_f} \sum_{f} {\overline \mu}_f^k \aleph(z) ,
\ 
{\widetilde \aleph}^{(k)}(\ell,z) = 
\frac{1}{n_f} \sum_{f} {\overline \mu}_f^k \aleph(\ell,z) ,
\label{not2}
\\
&& \ln^{\prime} \zeta(-n)  =  \frac{\zeta^{\prime}(-n)}{\zeta(-n)} \ ,
\label{not3}
\end{eqnarray}
where $\aleph(z)$ and $\aleph(\ell,z)$ are the aleph functions
defined in Refs.~\cite{Vuo} via digamma functions
and derivatives of the Riemann zeta functions.

Coefficients $A_j~(j = 1, ...7)$ and $A_j^{(\nu)}$, which appear in 
equations Eqs.~(\ref{mE2o1})-(\ref{lE12}) and (\ref{pEo1})-(\ref{pGo1}),
and later in Eqs.~(\ref{mE2})-(\ref{pE2}), 
are obtained from the ``matching
parameters'' $\alpha_{\rm E_i}~(i = 1, ... 7)$ in Ref.~\cite{Vuo} by
separating appropriately 
the parts proportional to $\ln \nu_c$ 
($\ln \bar \Lambda$ in Ref.~\cite{Vuo}), or proportional
to $1/\epsilon$, from the remaining ($\nu_c$- and 
$\epsilon$-independent) parts. They take the form
\begin{eqnarray}
&& A_1  =  \frac{\pi^2}{45} \left[ 8 + 3 n_f
\left( \frac{7}{4} + 30 {\widetilde \mu}_2 + 60 {\widetilde \mu}_4 \right) \right] \ ,
\label{AI1}
\\
&& A_2  =  - \frac{1}{6} \left[
1 + \frac{1}{12} n_f \left( 5 + 72 {\widetilde \mu}_2 + 144 {\widetilde \mu}_4 \right)
\right] \ ,
\label{AI2}
\\
&& A_4  =  \left[ 1 + \frac{1}{6} n_f \left( 1 + 12 {\widetilde \mu}_2 \right)
\right] \ ,
\label{AI4}
\\
&& A_5  =  2  
\left( - \ln 2 + \ln^{\prime} \zeta(-1) \right) +
\frac{1}{6} n_f (1 - 2 \ln 2) (1 + 12 {\widetilde \mu}_2) +
4 n_f {\widetilde \aleph}^{(0)}(1,z)  \ ,
\label{AI5}
\\
&& A_6  = {\bigg \{}
\frac{1}{18} \left[ 90 - 396 \ln 2 + 66 \gamma_{\rm E} (6 + n_f) 
+ n_f( 3 - 42 \ln 2) + 2 n_f^2 (1 + 2 \ln 2) \right]
\nonumber\\
&&~~ + \left[n_f {\widetilde \mu}_2 \left(
6 + 44 \gamma_{\rm E} - 44 \ln 2 + n_f \frac{4}{3} (1 +
2 \ln 2) \right) + 
\frac{1}{9} n_f (n_f + 6) {\widetilde \aleph}^{(0)}(z) +
n_f^2 \frac{4}{3} {\widetilde \aleph}^{(2)}(z) \right]
{\bigg \}} \ ,
\label{AI6}
\\
&& A_7  =  \left[ 22 (\gamma_{\rm E} - \ln 2) + 1 +
n_f \frac{4}{3} \ln 2 + n_f \frac{2}{3} {\widetilde \aleph}^{(0)}(z)
\right] \ ,
\label{AI7}
\end{eqnarray}
\begin{eqnarray}
&& A_5^{(\nu)}  =  2 A_4 \ ,
\label{A5L}
\\
&& A_6^{(\nu)}  =  22 + \frac{7}{3} n_f - \frac{2}{9} n^2_f + 4 (11 -
\frac{2}{3} n_f) n_f {\widetilde \mu}_2 \ ,
\label{A6L}
\\
&& A_7^{(\nu)}  =  22 - \frac{4}{3} n_f \ .
\label{A7L}
\end{eqnarray}
The most complicated coefficient is $A_3$, which emerges both in 
Eq.~(\ref{pEo1}) and later in Eq.~(\ref{pE2}) via the quantity $\kappa$ 
[Eq.~(\ref{kappa})]. It can be expressed
as
\begin{equation}
A_3  =  A_{3,1} + A_{3,2} - A_3^{(\nu)} \ln 2 \ ,
\label{AI3a}
\end{equation}
where
\begin{eqnarray}
A_{3,1} & = &
\frac{1}{18}
{\Bigg \{} 3^2 \left[ \frac{116}{5} + 4 \gamma_{\rm E}
- \frac{38}{3} \ln^{\prime} \zeta(-3) + 
\frac{220}{3} \ln^{\prime} \zeta(-1) \right]
\nonumber\\
&& + \frac{3}{2} n_f {\bigg [}
\frac{1121}{60} + 8 \gamma_{\rm E} + 2 (127 + 48 \gamma_{\rm E}) {\widetilde \mu}_2
- 644 {\widetilde \mu}_4 + \frac{268}{15} \ln^{\prime}\zeta(-3) +
\frac{4}{3} (11 + 156 {\widetilde \mu}_2) \ln^{\prime}\zeta(-1)
\nonumber\\
&& + 24 \left( 52 {\widetilde \aleph}^{(0)}(3,z) + 144 {\rm i} {\widetilde \aleph}^{(1)}(2,z)
+ 17 {\widetilde \aleph}^{(0)}(1,z) - 92 {\widetilde \aleph}^{(2)}(1,z) + 4 {\rm i} {\widetilde \aleph}^{(1)}(0,z)
\right)  
{\bigg ]}
\nonumber\\
&&+ \frac{2}{3} n_f {\bigg [}
\frac{3}{4} ( 35 + 472 {\widetilde \mu}_2 + 1328 {\widetilde \mu}_4 ) - 
24 (1 - 4 {\widetilde \mu}_2) \ln^{\prime}\zeta(-1)
\nonumber\\
&&- 144 \left( 12 {\rm i} {\widetilde \aleph}^{(1)}(2,z) - 2 {\widetilde \aleph}^{(0)}(1,z)
- 16 {\widetilde \aleph}^{(2)}(1,z) - {\rm i} {\widetilde \aleph}^{(1)}(0,z)
- {\rm i} 4 {\widetilde \aleph}^{(3)}(0,z) \right)
{\bigg ]}
\nonumber\\
&&+ \frac{1}{4} n_f^2 {\bigg [}
\frac{1}{3} + 4 \gamma_{\rm E} + 8 (7 + 12 \gamma_{\rm E}) {\widetilde \mu}_2 + 112 {\widetilde \mu}_4
- \frac{64}{15} \ln^{\prime}\zeta(-3)
- \frac{32}{3} (1 + 12 {\widetilde \mu}_2) \ln^{\prime}\zeta(-1)
\nonumber\\
&& - 96 \left( 8 {\widetilde \aleph}^{(0)}(3,z) + {\rm i} 12 {\widetilde \aleph}^{(1)}(2,z)
- 2 {\widetilde \aleph}^{(0)}(1,z) - 4 {\widetilde \aleph}^{(2)}(1,z) - {\rm i} {\widetilde \aleph}^{(1)}(0,z)
\right)
{\bigg ]}
{\Bigg \}} \ ,
\label{AI31}
\end{eqnarray}
\begin{eqnarray}
A_{3,2} & = &
4 \sum_{f,g}
{\bigg [}
2 (1 + \gamma_{\rm E}) {\overline \mu}_f^2 {\overline \mu}_g^2 
- \aleph(3,z_f+z_g) - \aleph(3,z_f+z_g^{\ast}) -
{\rm i} 4 {\overline \mu}_f \left( \aleph(2,z_f+z_g) + \aleph(2,z_f+z_g^{\ast})
 \right)
\nonumber\\
&& + 4 {\overline \mu}_g^2 \aleph(1,z_f) + ({\overline \mu}_f + {\overline \mu}_g)^2 \aleph(1,z_f+z_g)
+ ({\overline \mu}_f - {\overline \mu}_g)^2 \aleph(1,z_f+z_g^{\ast})
+ {\rm i} 4 {\overline \mu}_g^2 {\overline \mu}_f \aleph(0,z_f)
{\bigg ]} \ ,
\label{AI32}
\end{eqnarray}
and
\begin{eqnarray}
A^{\rm (\nu)}_3 & = &
\left[
\left( \frac{97}{3} + \frac{169}{36} n_f + \frac{5}{54} n_f^2 \right)
+ {\widetilde \mu}_2 n_f \left( 50 + \frac{4}{3} n_f \right)
+ {\widetilde \mu}_4 n_f \left( - 44 + \frac{8}{3} n_f \right)
\right] \ .
\label{A3L}
\end{eqnarray}
The constants $\alpha_{\rm G}$, $\alpha_{\rm M1}$ and $\alpha_{\rm M2}$
were obtained in Ref.~\cite{KLRS1}:
\begin{equation}
\alpha_{\rm G} = \frac{43}{96} - \frac{157}{6144} \pi^2
\approx 0.195715, \quad
\alpha_{\rm M1} = \frac{43}{32} - \frac{491}{6144} \pi^2
\approx 0.555017, \quad
\alpha_{\rm M2} = - \frac{4}{3} \ ,
\label{alphas}
\end{equation}
and $\beta_{\rm M1}$ in Ref.~\cite{KLRS2}:
\begin{equation}
\beta_{\rm M1}  \approx -1.391512 \ .
\label{beM}
\end{equation}

\section{Beta functions of the Borel-Pad\'e type}
\label{app2}

In this Appendix we present various resummations of the QCD $\beta$
functions as functions of $x = a(Q^2) = \alpha_s(Q^2)/\pi$,
in the ${\overline {\rm MS}}$ scheme. Further, the corresponding running
of $a(\nu^2)$ as function of $x=\nu^2$ (in ${\rm GeV}^2$)
is given, in the various cases, always normalized to
$a(m_{\tau}^2) = 0.334/\pi = 0.106316$, cf.~Eq.~(\ref{alsmtau}).
We will denote the squared RScl $\nu^2$ here as $Q^2 (\equiv - q^2 > 0)$
to emphasize the space-like character of the corresponding
four-vector $q$.

The four-loop renormalization group equation (RGE) for 
$a(Q^2) \equiv \alpha_s(Q^2)/\pi$ is:
\begin{equation}
Q^2 \frac{d a(Q^2)}{Q^2} = - \beta_0 a^2(Q^2)
\left[ 1 + c_1 a(Q^2) + c_2 a^2(Q^2) + c_3 a^3(Q^2) \right]
\ ,
\label{RGETPS}
\end{equation}
where $c_j \equiv \beta_j/\beta_0$ ($j \geq 1$).
The one- and two-loop coefficients $\beta_0$ and $\beta_1$
\cite{1lbe,2lbe} are scheme independent; in ${\overline {\rm MS}}$
scheme \cite{'tHooft:1973mm} the
three- and four-loop coefficients $\beta_2$ and $\beta_3$
were obtained in Refs.~\cite{3lbebMS,4lbebMS}, respectively
\begin{eqnarray}
\beta_0 &=& \frac{1}{4} 
\left( 11 - \frac{2}{3} n_f \right) \ ,
\qquad
\beta_1 =  \frac{1}{16}
\left( 102 - \frac{38}{3} n_f \right) \ ,
\label{be01} 
\\
\beta_2 & = & \frac{1}{64}
\left( \frac{2857}{2} - \frac{5033}{18} n_f 
+ \frac{325}{54} n_f^2 \right) \ ,
\label{be2} 
\\
\beta_3 & = & \frac{1}{256}
( 29243.0 - 6946.30 \ n_f 
+ 405.089 \ n_f^2 + 1.49931 \ n_f^3 ) \ ,
\label{be3} 
\end{eqnarray}
and $n_f$ is the active number of quark flavors.

\begin{figure}[htb]
\begin{minipage}[b]{.49\linewidth}
\centering\epsfig{file=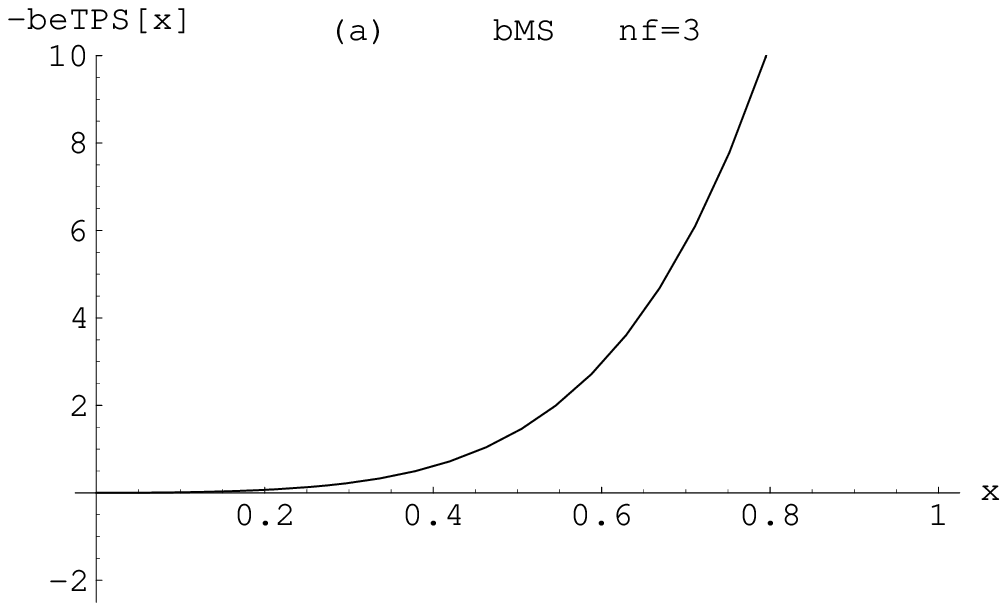,width=\linewidth}
\end{minipage}
\begin{minipage}[b]{.49\linewidth}
 \centering\epsfig{file=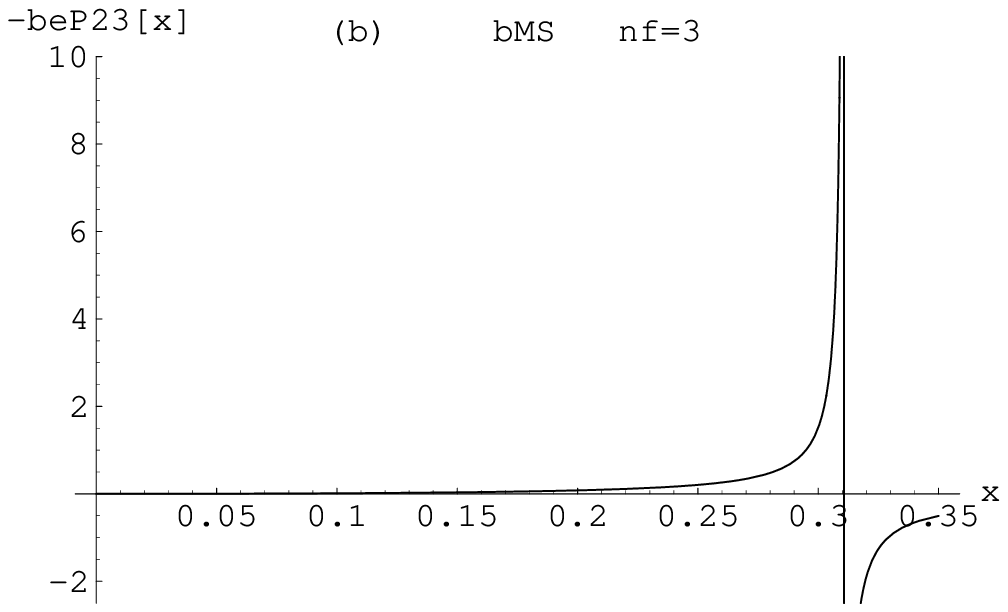,width=\linewidth}
\end{minipage}
\caption{\footnotesize ${\overline {\rm MS}}$ 
beta functions 
$-\beta(x)/\beta_0$, for $n_f=3$,
whose TPS is ${\rm TPS}(x)=1 + c_1 x + c_2 x^2 + c_3 x^3$:
(a) TPS(x), and (b) Pad\'e P[2/3](x) cases;
$x$ here stands for $a(Q^2)$.}
\label{beTPSP23nf3}
\end{figure}

The solution of the RGE (\ref{RGETPS}), at low
Euclidean energies Q (with either $n_f=3$ or $n_f=2$) has
the known unphysical Landau singularities, i.e.,
singularities of $a(Q^2)$ for $Q^2 \leq Q^2_{\rm pole}$.
For example, if we choose the realistic value
of $\alpha_s(Q^2=m^2_{\tau}) = 0.334$, the 
singularity in $a(Q^2)$
appears already at $Q^2_{\rm pole} \approx 
0.66^2 \ {\rm GeV}^2 \approx 0.44 \ {\rm GeV}^2$
when $n_f=3$, and
 $Q^2_{\rm pole} \approx 
0.75^2 \ {\rm GeV}^2 \approx 0.57 \ {\rm GeV}^2$
when $n_f=2$
-- cf.~Figs.~\ref{asbeTPSP23nf3}(a) and
\ref{asbeTPSP23nf2}(a).
The main reason for this unphysical behavior
of $a(Q^2)$ is the truncated perturbation
series (TPS) form of the beta function 
$\beta(x=a)$ -- the right-hand side of RGE (\ref{RGETPS}).
Such a form of $\beta(x)$ has its origin in the 
perturbative approach (powers of $x=a$). The
TPS $|\beta(x)|$ grows out of control when
$x=a$ increases -- cf.~Figs.~\ref{beTPSP23nf3}(a)
and \ref{beTPSP23nf2}(a).
This leads to the appearance
of the nonphysical singularities in $a(Q^2)$ 
at low positive $Q^2 \leq Q^2_{\rm pole}$.

These singularities prevent us from
using, in the traditional perturbative QCD (pQCD),
the coupling at squared energies $Q^2 \alt Q^2_{\rm pole}$.
The problem can be avoided by certain resummations
of the TPS $\beta$ function, i.e, by finding such
a $\beta(a)$ function whose Taylor expansion
around $a=0$ up to $\sim a^5$ reproduces the 
TPS $\beta(a)$ of Eq.~(\ref{RGETPS}) and, at the same time,
$\beta(a)$ remains more under control when $a$ increases.

One possibility is to construct diagonal or
near-to-diagonal Pad\'e approximants based on the
TPS $\beta(a)$. For example, the Pad\'e
\begin{equation}
{\rm P}[2/3]_{\beta}(a) = - \beta_0 \frac{a^2}{ \left[ 1 - c_1 a +
(c_1^2 - c_2) a^2 + (- c_1^3 + 2 c_1 c_2 - c_3) a^3  \right]} 
\label{beP23}
\end{equation}
gives us an expression which, up to
$a \approx 0.3$, behaves well
-- cf.~Figs.~\ref{beTPSP23nf3}(b) and
 \ref{beTPSP23nf2}(b). However, around $a \approx 0.3$,
this $\beta(a)$ goes abruptly out of control, because
the Pad\'e expression has a pole there.
The corresponding running coupling $a(x=Q^2)$
achieves singularity already at   
$Q_{\rm pole}^2 \approx 0.81^2 \ {\rm GeV^2}
\approx 0.65 \ {\rm GeV^2}$
for $n_f=3$, and 
$Q_{\rm pole}^2 \approx 0.92^2 \ {\rm GeV^2}
\approx 0.85 \ {\rm GeV^2}$
for $n_f=2$ -- cf.~Figs.~\ref{asbeTPSP23nf3}(b)
and \ref{asbeTPSP23nf2}(b). In contrast to the
TPS $\beta(a)$ case, however, $a(Q^2)$ seems to
be well under control now for virtually all
$Q^2$ larger than $Q_{\rm pole}^2$.

\begin{figure}[htb]
\begin{minipage}[b]{.49\linewidth}
\centering\epsfig{file=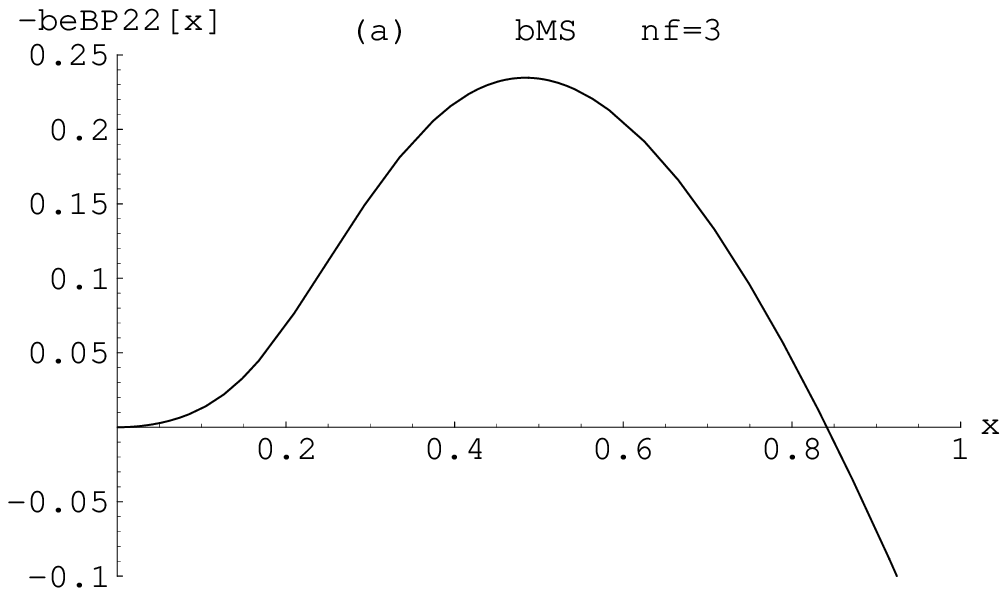,width=\linewidth}
\end{minipage}
\begin{minipage}[b]{.49\linewidth}
 \centering\epsfig{file=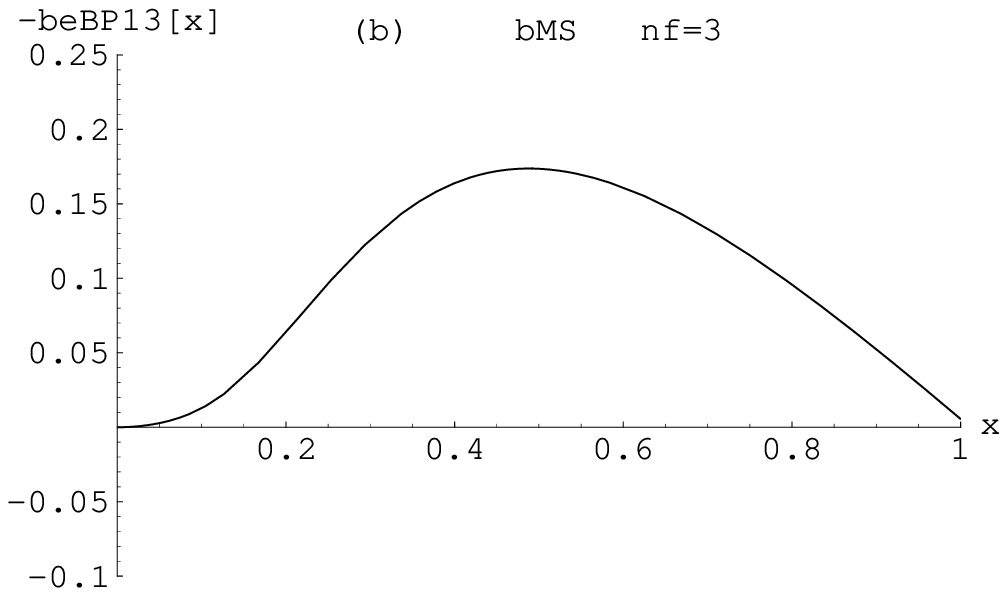,width=\linewidth}
\end{minipage}
\caption{\footnotesize ${\overline {\rm MS}}$
beta functions 
$-\beta(x)/\beta_0$, for $n_f=3$,
whose expansion is $(1 + c_1 x + c_2 x^2 + c_3 x^3)$:
(a) Borel-Pad\'e BP[2/2](x), and (b) BP[1/3](x) cases;
$x$ here stands for $a(Q^2)$.}
\label{beBPnf3}
\end{figure}

\begin{figure}[htb]
\begin{minipage}[b]{.49\linewidth}
\centering\epsfig{file=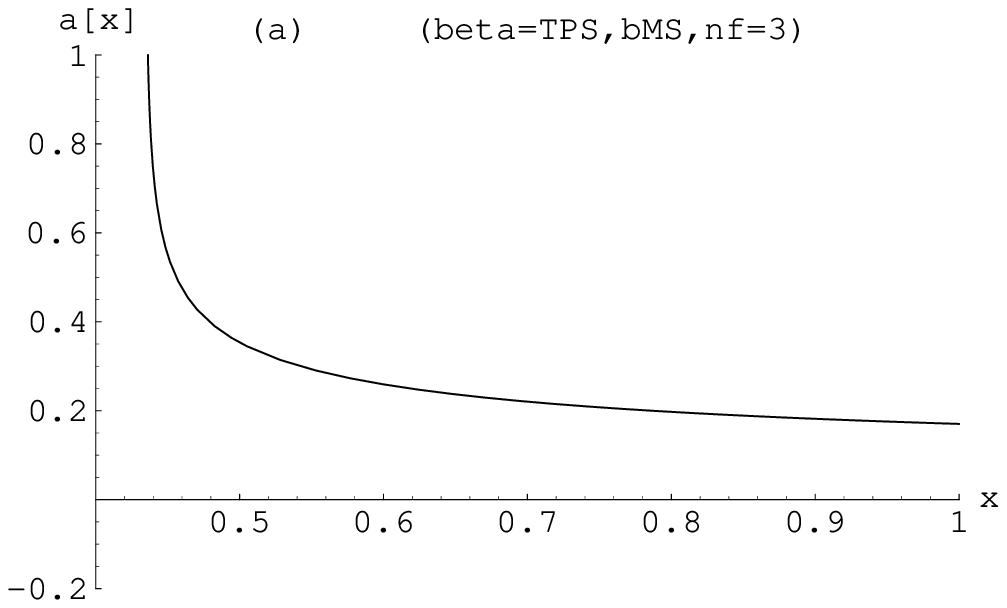,width=\linewidth}
\end{minipage}
\begin{minipage}[b]{.49\linewidth}
 \centering\epsfig{file=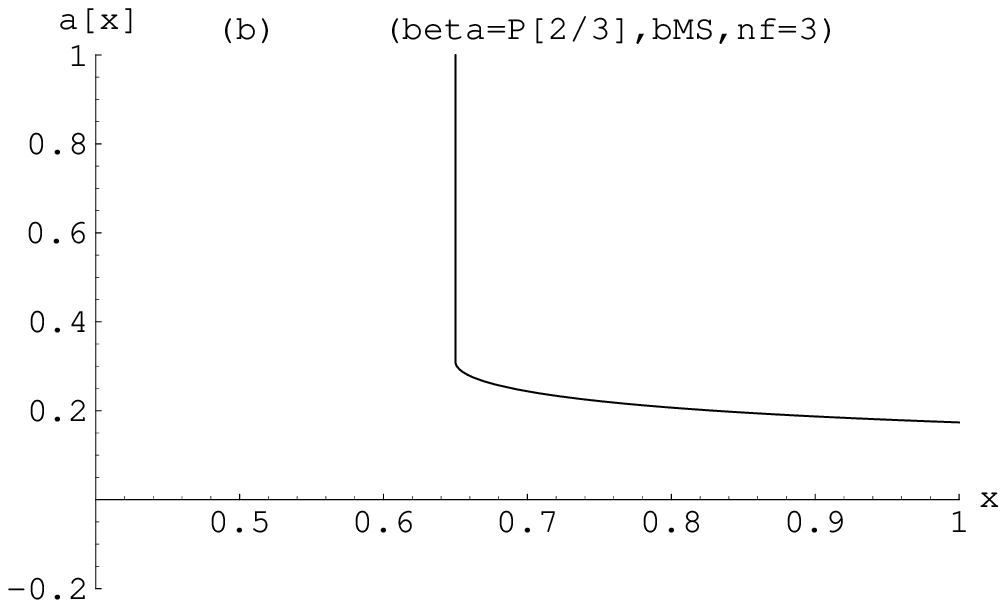,width=\linewidth}
\end{minipage}
\caption{\footnotesize ${\overline {\rm MS}}$ running 
of $a(x=Q^2)$, for $n_f=3$,
when the beta functions are (a) TPS, and (b) Pad\'e P[2/3].}
\label{asbeTPSP23nf3}
\end{figure}

\begin{figure}[htb]
\begin{minipage}[b]{.49\linewidth}
\centering\epsfig{file=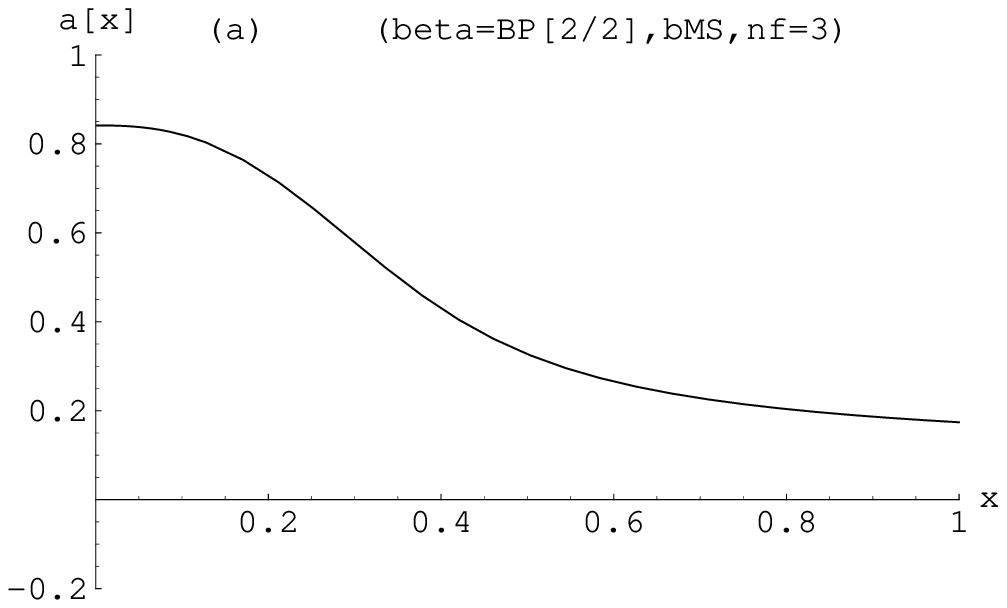,width=\linewidth}
\end{minipage}
\begin{minipage}[b]{.49\linewidth}
 \centering\epsfig{file=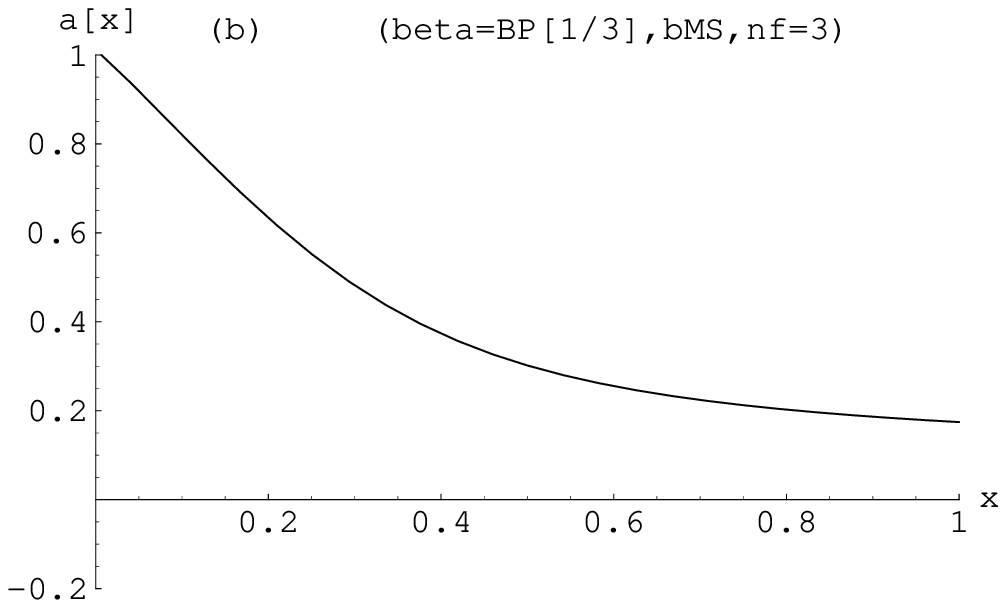,width=\linewidth}
\end{minipage}
\caption{\footnotesize ${\overline {\rm MS}}$ running 
of $a(x=Q^2)$, for $n_f=3$,
when the beta functions are (a) Borel-Pad\'e BP[2/2], 
and (b) BP[1/3].}
\label{asbeBPnf3}
\end{figure}

\begin{figure}[htb]
\begin{minipage}[b]{.49\linewidth}
\centering\epsfig{file=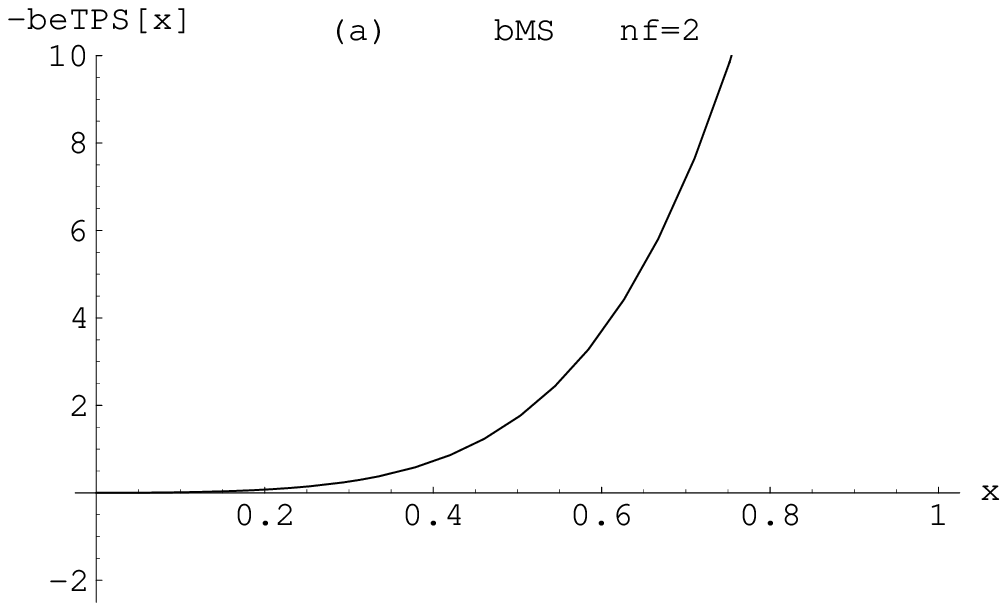,width=\linewidth}
\end{minipage}
\begin{minipage}[b]{.49\linewidth}
 \centering\epsfig{file=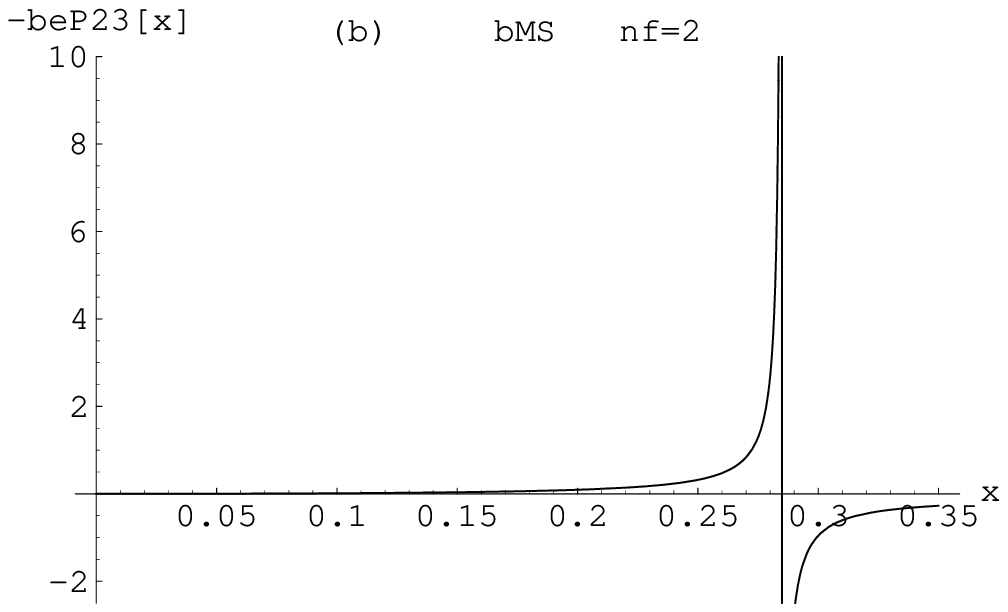,width=\linewidth}
\end{minipage}
\caption{\footnotesize ${\overline {\rm MS}}$ 
beta functions 
$-\beta(x)/\beta_0$, for $n_f=2$,
whose TPS is ${\rm TPS}(x)=1 + c_1 x + c_2 x^2 + c_3 x^3$:
(a) TPS(x), and (b) Pad\'e P[2/3](x) cases;
$x$ here stands for $a(Q^2)$.}
\label{beTPSP23nf2}
\end{figure}

\begin{figure}[htb]
\begin{minipage}[b]{.49\linewidth}
\centering\epsfig{file=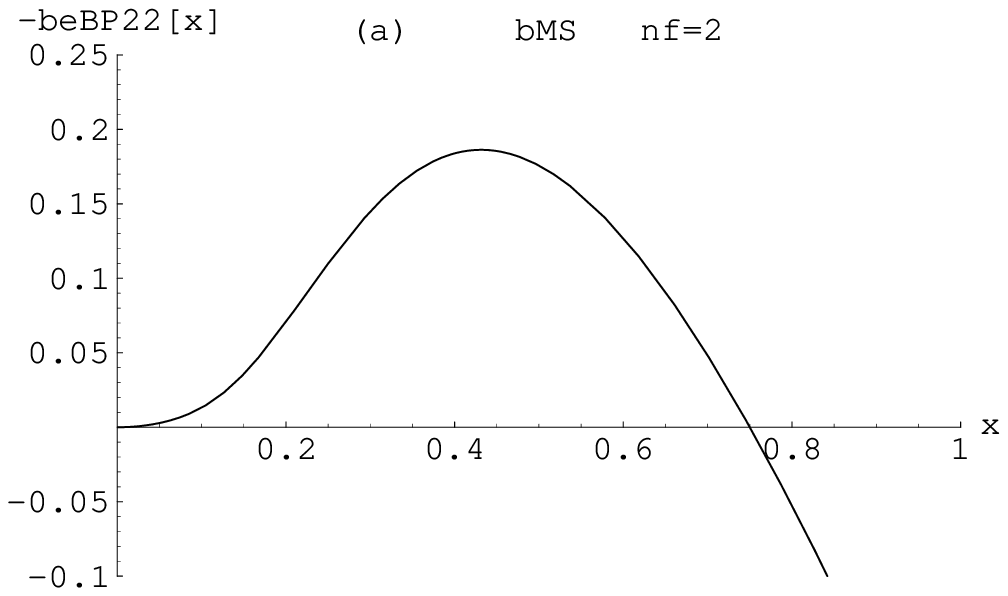,width=\linewidth}
\end{minipage}
\begin{minipage}[b]{.49\linewidth}
 \centering\epsfig{file=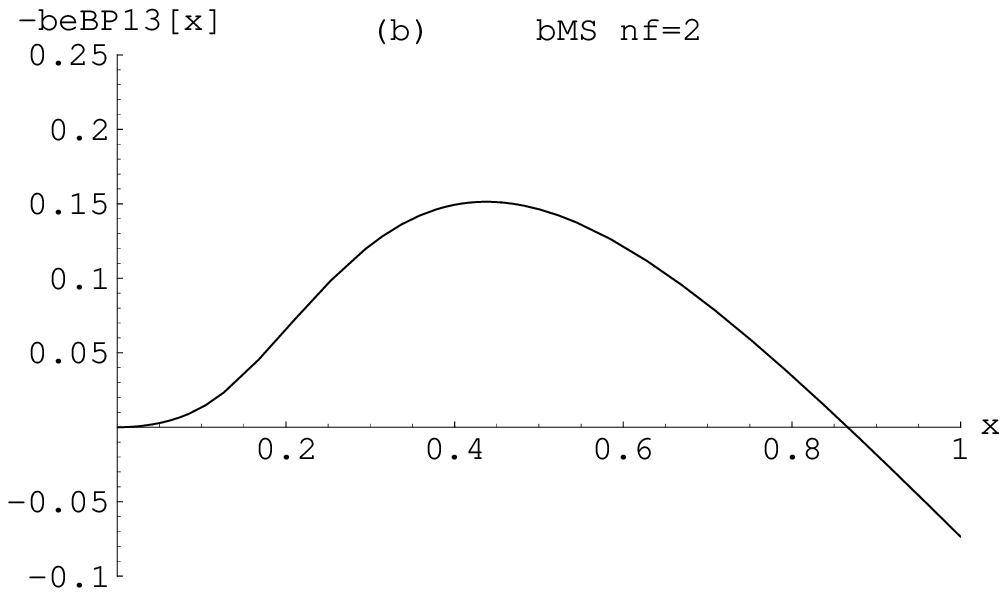,width=\linewidth}
\end{minipage}
\caption{\footnotesize ${\overline {\rm MS}}$
beta functions 
$-\beta(x)/\beta_0$, for $n_f=2$,
whose expansion is $(1 + c_1 x + c_2 x^2 + c_3 x^3)$:
(a) Borel-Pad\'e BP[2/2](x), and (b) BP[1/3](x) cases;
$x$ here stands for $a(Q^2)$.}
\label{beBPnf2}
\end{figure}

\begin{figure}[htb]
\begin{minipage}[b]{.49\linewidth}
\centering\epsfig{file=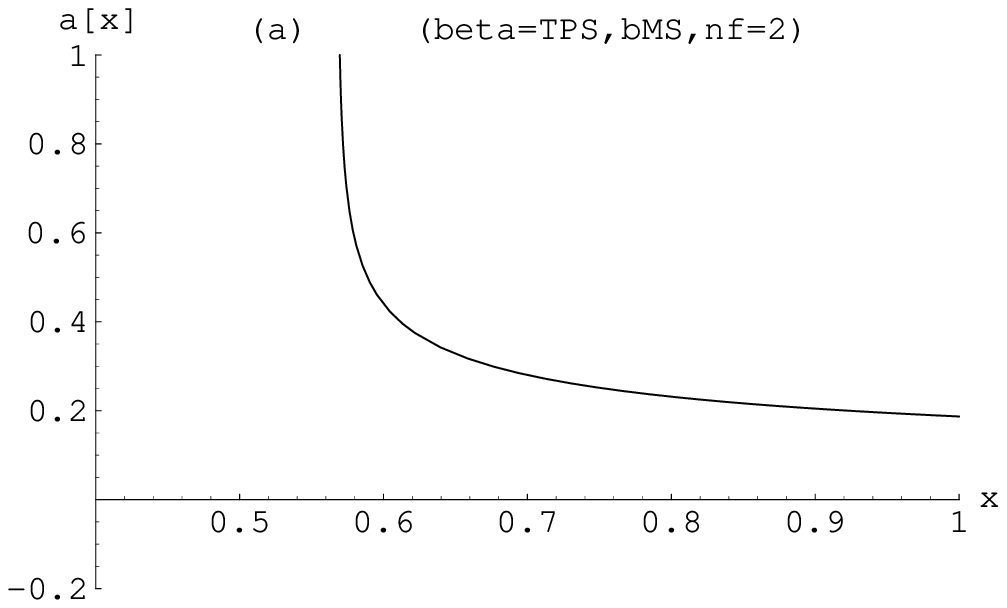,width=\linewidth}
\end{minipage}
\begin{minipage}[b]{.49\linewidth}
 \centering\epsfig{file=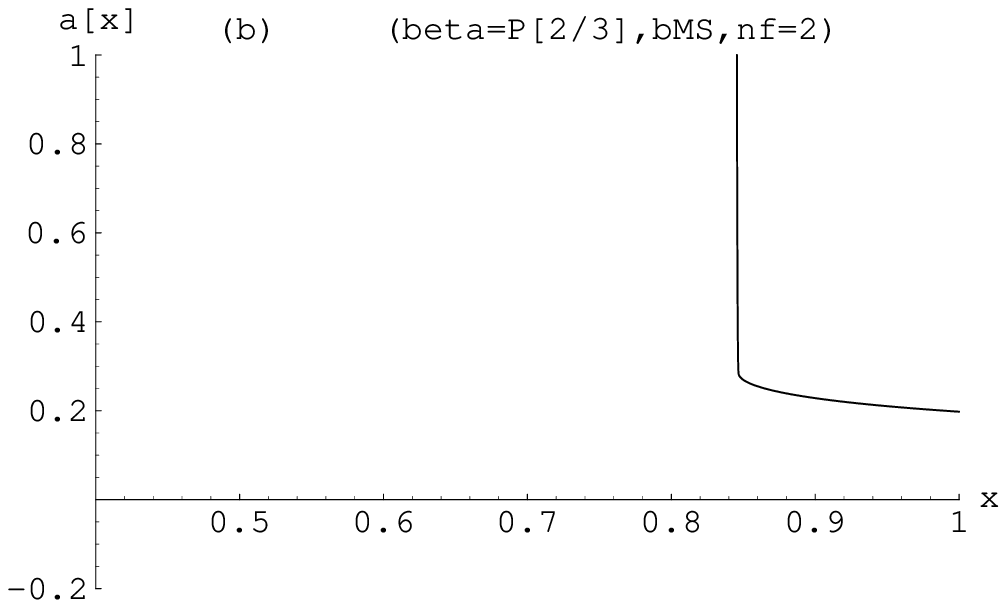,width=\linewidth}
\end{minipage}
\caption{\footnotesize ${\overline {\rm MS}}$ running 
of $a(x=Q^2)$, for $n_f=2$,
when the beta functions are (a) TPS, and (b) Pad\'e P[2/3].}
\label{asbeTPSP23nf2}
\end{figure}

\begin{figure}[htb]
\begin{minipage}[b]{.49\linewidth}
\centering\epsfig{file=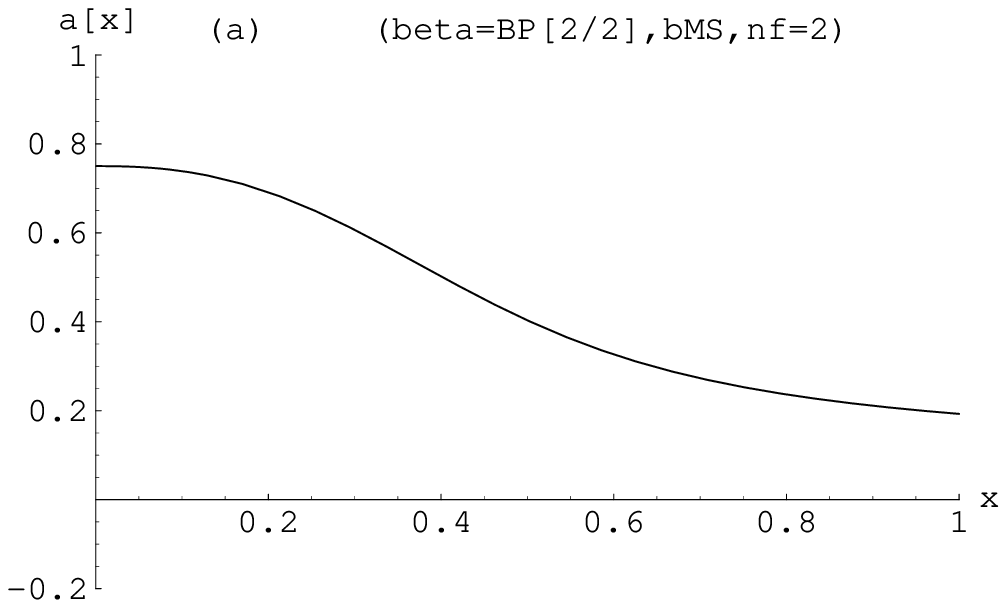,width=\linewidth}
\end{minipage}
\begin{minipage}[b]{.49\linewidth}
 \centering\epsfig{file=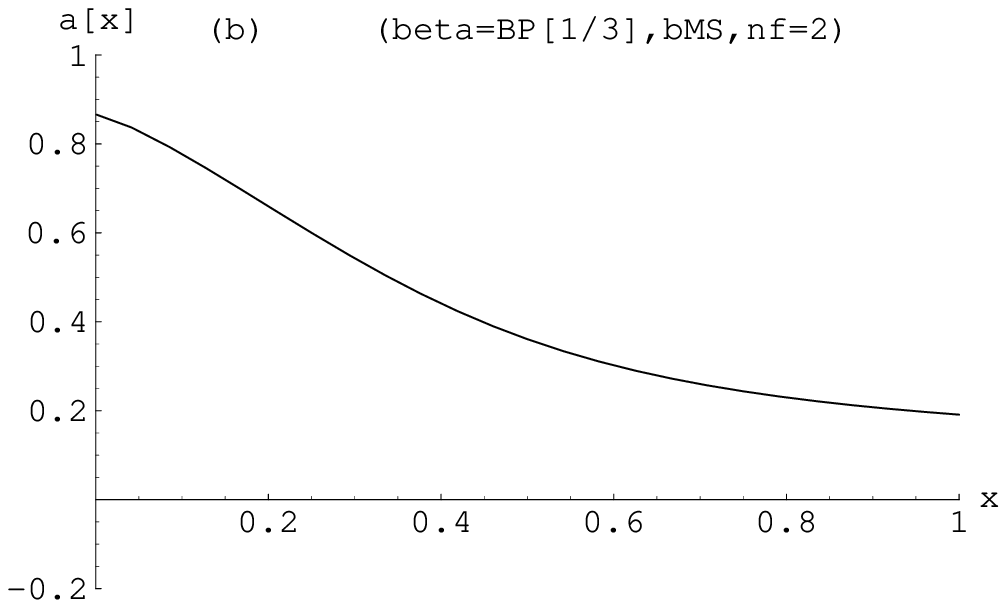,width=\linewidth}
\end{minipage}
\caption{\footnotesize ${\overline {\rm MS}}$ running 
of $a(x=Q^2)$, for $n_f=2$,
when the beta functions are (a) Borel-Pad\'e BP[2/2], 
and (b) BP[1/3].}
\label{asbeBPnf2}
\end{figure}

Another possibility, which avoids the aforementioned
pole problem of the Pad\'e $\beta(x=a)$, would go
in the direction of resumming first the
Borel transform ${\rm B}_{\beta}(y)$ 
(the latter has in general significantly
weaker singularities than $\beta$),
and then applying the inverse transformation
via a Borel integration. For example, we can
try to apply diagonal or close-to-diagonal
Pad\'e resummation to ${\rm B}_{\beta}(y)$.
The Borel transform is
\begin{eqnarray}
{\rm B}_{\beta}(y) & = & -\beta_0 \left( 
\frac{y}{1!} + c_1 \frac{y^2}{2!} +
c_2 \frac{y^3}{3!} + c_3 \frac{y^4}{4!} + \cdots
\right) \ ,
\label{BbetaTPS}
\end{eqnarray}
and the Pad\'e P[2/2] and P[1/3] resummations
of the above TPS are
\begin{eqnarray}
{\rm P[2/2]_B}(y) & = & - \beta_0
\frac{y + r_2 y^2}{1 + t_1 y + t_2 y^2} \ ,
\label{B22}
\\ 
{\rm P[1/3]_B}(y) & = & -\beta_0
\frac{y}{1 + s_1 y + s_2 y^2 + s_3 y^3} \ ,
\label{B13}
\end{eqnarray}
where the coefficients $r_j, t_j, s_j$ are unique functions
of $c_k$'s such that reexpansion of (\ref{B22})
and (\ref{B13}) reproduces expansion (\ref{BbetaTPS})
up to (and including) $\sim y^4$ term:
\begin{eqnarray}
r_2 & = & (1/2) (3 c_1^3 - 4 c_1 c_2 + c_3)/\xi \ ,
\nonumber\\
t_1 & = & (1/2) (- 2 c_1 c_2 + c_3)/\xi \ ,
\qquad
t_2 = (1/12) (4 c_2^2 - 3 c_1 c_3)/\xi \ ,
\label{coeffrt}
\\
s_1 & = & (1/2) c_1 \ , 
\qquad
s_2 = (1/12) (3 c_1^2 - 2 c_2) \ ,
\qquad
s_3 = (1/24) ( - 3 c_1^3 + 4 c_1 c_2 - c_3 ) \ ,
\label{coeffs}
\end{eqnarray}
and we used the notation $\xi = (3 c_1^2 - 2 c_2)$.
However, the inverse Borel transformation 
\begin{equation}
\beta(x) = \int_{0}^{\infty} 
dy \ \exp(-y/x) {\rm B}_{\beta}(y) 
\label{Borel1}
\end{equation}
cannot be constructed by inserting here directly
the Pad\'e expressions  (\ref{B22}) or (\ref{B13}) for the integrand. 
This is so because the latter
expressions have poles on the positive axis:
${\rm P[2/2]_B}(y)$ at $y_{\rm p.}\approx 1.13, 1.01$
for $n_f=3,2$, respectively;
${\rm P[1/3]_B}(y)$ at $y_{\rm p.}\approx 0.94, 0.88$
for $n_f=3,2$, respectively.
These ``infrared renormalon'' singularities
imply ambiguities in the integration:
$\delta \beta(a) \sim \exp(- y_{\rm p.}/a)$
[$\sim \delta ( Q^2 d a(Q^2)/d Q^2 ) \sim \delta a(Q^2)$].
This implies an ambiguity in the coupling $a(Q^2)$:
$\delta a(Q^2) \sim 
({\Lambda^2_{\rm {\overline {MS}}}}/Q^2 )^{\beta_0 y_{\rm p.}}
\equiv ({\Lambda^2_{\rm {\overline {MS}}}}/Q^2 )^{\eta}$.
Numerically, for ${\rm P[2/2]_B}$,
$\eta \approx 2.55, 2.44$ for $n_f=3,2$, respectively;
for ${\rm P[1/3]_B}$,
$\eta \approx 2.11, 2.13$ for $n_f=3,2$, respectively.
We can fix the above ambiguity by choosing
a specific recipe for the Borel integration over the
pole $y_{\rm p.}$. We will choose the Principal
Value (PV) prescription
\begin{equation}
{\rm BP}[i/j]_{\beta}(x) = {\rm Re} 
\int_{\pm i \varepsilon}^{\infty \pm i \varepsilon} 
dy \ \exp(-y/x) {\rm P}[i/j]_{\rm B} (y) \ ,
\label{Borel2}
\end{equation}
where $[i/j] = [2/2]$ or $[1/3]$.
Numerically, this is difficult to implement,
as $\varepsilon \to +0$ and we approach the
pole $y_{\rm p.}$ down to the distance ${\varepsilon}$
during the integration. However, we can use the Cauchy
theorem, and the fact that the Borel transforms
(\ref{B22}) and (\ref{B13}) do not have any
poles in the complex semiplane
${\rm Re}(y) \geq 0$ except the aforementioned
$y_{\rm p.} > 0$. This allows us to
avoid the vicinity of the pole,
for example by integrating along a ray
$y = r \exp(- i \phi)$, where $\phi$ is
any small but finite positive fixed angle (cf.~Refs.~\cite{Cvetic:2001sn,GRp})
\begin{eqnarray}
{\rm BP}[i/j]_{\beta}(x) & = & {\rm Re} 
{\Big \{}
\exp(- i \phi )
\int_{r=0}^{\infty} dr \exp(- y/x ) {\rm P}[i/j]_{\rm B} (y)
\Big|_{y=r \exp(-i \phi)} {\Big \}} \ .
\label{Borel3}
\end{eqnarray} 
This approach, which is numerically stable, gives us
for the Borel-Pad\'e(BP)-resummed $\beta(x)$-functions
values which are surprisingly non-singular and
achieve at $x \equiv a \approx 1.$ value zero
(infrared fixed point) -- cf.~Figs.~\ref{beBPnf3}
and \ref{beBPnf2}. Integration of the RGE with these
BP-resummed four-loop ${\rm {\overline {MS}}}$
$\beta$ functions, with the phenomenologically
acceptable initial condition
$\alpha_s(m^2_{\tau}) = 0.334$ [Eq.~(\ref{alsmtau})], gives us
$a(x=Q^2)$ running couplings which are
presented in Figs.~\ref{asbeBPnf3} and
\ref{asbeBPnf2}. Both choices BP[2/2]
and BP[1/3] give similar behavior for
$a(Q^2)$. There are no Landau singularities present
any more, and the coupling is analytic in the
sense that there are no unphysical singularities 
on the space-like axis of the squared momenta
$q^2 \equiv - Q^2$. Furthermore, due to
the aforementioned zero of the BP $\beta$ functions,
the coupling $a(Q^2)$ remains finite down to
$Q^2=0$ where it has a value $\approx 1$.
The obtained ``analytized'' coupling $a(Q^2)$
probably represents a version of analytic QCD.
Therefore, the skeleton-motivated method of Refs.~\cite{GCCV}
can be applied as a alternative way of evaluating the
low-energy QCD observables.

\clearpage
\newpage

\clearpage
\newpage

\clearpage
\newpage

\clearpage
\newpage

\begin{thebibliography}{99}

\bibitem{Fodor} 
  Z.~Fodor and S.~D.~Katz,
  Phys.\ Lett.\ B {\bf 534}, 87 (2002)
  [hep-lat/0104001];
  JHEP {\bf 0203}, 014 (2002)
  [hep-lat/0106002].

\bibitem{Allton} 
  C.~R.~Allton {\it et al.},
  Phys.\ Rev.\ D {\bf 66}, 074507 (2002)
  [hep-lat/0204010];
  C.~R.~Allton {\it et al.},
  Nucl.\ Phys.\ Proc.\ Suppl.\  {\bf 119}, 538 (2003)
  [hep-lat/0209012].

\bibitem{Forcrand} 
  P.~de Forcrand and O.~Philipsen,
  Nucl.\ Phys.\ B {\bf 642}, 290 (2002)
  [hep-lat/0205016];
  M.~D'Elia and M.~P.~Lombardo,
  Phys.\ Rev.\ D {\bf 67}, 014505 (2003)
  [hep-lat/0209146].

\bibitem{Krat} 
  S.~Kratochvila and P.~de Forcrand,
  Nucl.\ Phys.\ Proc.\ Suppl.\  {\bf 140}, 514 (2005)
  [hep-lat/0409072].

\bibitem{KLRS1}
K.~Kajantie, M.~Laine, K.~Rummukainen and Y.~Schr\"oder,
Phys.\ Rev.\ D {\bf 67}, 105008 (2003)
[hep-ph/0211321].

\bibitem{Vuo} A.~Vuorinen,
  Phys.\ Rev.\ D {\bf 68}, 054017 (2003)
  [hep-ph/0305183];
  hep-ph/0402242;
  Phys.\ Rev.\ D {\bf 67}, 074032 (2003)
  [hep-ph/0212283].

\bibitem{Padebook}
George A. Baker, Jr. and Peter Graves-Morris,
{\it Pad\'e Approximants\/}, 2nd edition,
(Encyclopedia of Mathematics and Its Applications, Vol.~59),
edited by Gian-Carlo Rota
(Cambridge University Press, 1996).


\bibitem{Gardi} 
E.~Gardi,
Phys.\ Rev.\ D {\bf 56}, 68 (1997)
[hep-ph/9611453].

\bibitem{Cvetic} 
  G.~Cveti\v{c},
  Nucl.\ Phys.\ B {\bf 517}, 506 (1998)
  [hep-ph/9711406];
  Phys.\ Rev.\ D {\bf 57}, R3209 (1998)
  [hep-ph/9711487];
  Nucl.\ Phys.\ Proc.\ Suppl.\  {\bf 74}, 333 (1999)
  [hep-ph/9808273];
  G.~Cveti\v{c} and R.~K\"ogerler,
  Nucl.\ Phys.\ B {\bf 522}, 396 (1998)
  [hep-ph/9802248].



\bibitem{CK2}
 G.~Cveti\v{c},
  Phys.\ Lett.\ B {\bf 486}, 100 (2000)
  [hep-ph/0003123];
  G.~Cveti\v{c} and R.~K\"ogerler,
  Phys.\ Rev.\ D {\bf 63}, 056013 (2001)
  [hep-ph/0006098].

\bibitem{GRp}
  G.~Cveti\v{c} and R.~K\"ogerler,
  Phys.\ Rev.\ D {\bf 70}, 114016 (2004)
  [hep-ph/0406028];
  Phys.\ Rev.\ D {\bf 66}, 105009 (2002)
  [hep-ph/0207291].

\bibitem{Gross} 
  D.~J.~Gross, R.~D.~Pisarski and L.~G.~Yaffe,
  Rev.\ Mod.\ Phys.\  {\bf 53}, 43 (1981);
  T.~Appelquist and R.~D.~Pisarski,
  Phys.\ Rev.\ D {\bf 23}, 2305 (1981).

\bibitem{Braaten:1996ju}
E.~Braaten and A.~Nieto,
Phys.\ Rev.\ Lett.\  {\bf 76}, 1417 (1996)
[hep-ph/9508406];
Phys.\ Rev.\ D {\bf 53}, 3421 (1996)
[hep-ph/9510408].


\bibitem{HLP}
  A.~Hart, M.~Laine and O.~Philipsen,
  Nucl.\ Phys.\ B {\bf 586}, 443 (2000)
  [hep-ph/0004060].

\bibitem{DiRenzo:2006nh}
  F.~Di Renzo, M.~Laine, V.~Miccio, Y.~Schr\"oder and C.~Torrero,
  JHEP {\bf 0607}, 026 (2006)
  [hep-ph/0605042].

\bibitem{Nadkarni:1988fh}
S.~Nadkarni,
Phys.\ Rev.\ D {\bf 38}, 3287 (1988).

\bibitem{Landsman:1989be}
N.~P.~Landsman,
Nucl.\ Phys.\ B {\bf 322}, 498 (1989).

\bibitem{Geshkenbein:2001mn}
B.~V.~Geshkenbein, B.~L.~Ioffe and K.~N.~Zyablyuk,
Phys.\ Rev.\ D {\bf 64}, 093009 (2001)
[hep-ph/0104048].

\bibitem{Cvetic:2001sn}
G.~Cveti\v{c} and T.~Lee,
Phys.\ Rev.\ D {\bf 64}, 014030 (2001)
[hep-ph/0101297];
G.~Cveti\v{c}, C.~Dib, T.~Lee and I.~Schmidt,
Phys.\ Rev.\ D {\bf 64}, 093016 (2001)
[hep-ph/0106024].

\bibitem{Allton:2003vx}
C.~R.~Allton, S.~Ejiri, S.~J.~Hands, O.~Kaczmarek, F.~Karsch, 
E.~Laermann and C.~Schmidt,
  Phys.\ Rev.\ D {\bf 68}, 014507 (2003)
  [hep-lat/0305007].

\bibitem{Allton:2005gk}
  C.~R.~Allton {\it et al.},
  Phys.\ Rev.\ D {\bf 71}, 054508 (2005)
  [hep-lat/0501030].

\bibitem{Fodor:2002km}
  Z.~Fodor, S.~D.~Katz and K.~K.~Szabo,
  Phys.\ Lett.\ B {\bf 568}, 73 (2003)
  [hep-lat/0208078].

\bibitem{Csikor:2004ik}
  F.~Csikor, G.~I.~Egri, Z.~Fodor, S.~D.~Katz, K.~K.~Szabo and A.~I.~Toth,
  JHEP {\bf 0405}, 046 (2004)
  [hep-lat/0401016].

\bibitem{math}
{\it Mathematica\/} 5.2, Wolfram Research, Inc. 

\bibitem{Gliozzi:2007jh}
  F.~Gliozzi,
hep-lat/0701020.

\bibitem{Bernard:2002yd}
  C.~Bernard {\it et al.}  [MILC Collaboration],
  Nucl.\ Phys.\ Proc.\ Suppl.\  {\bf 119}, 523 (2003)
  [hep-lat/0209079].

\bibitem{Blaizot:2001vr}
  J.~P.~Blaizot, E.~Iancu and A.~Rebhan,
  Phys.\ Lett.\ B {\bf 523}, 143 (2001)
  [hep-ph/0110369].
 
\bibitem{GG}
  R.~V.~Gavai, S.~Gupta and P.~Majumdar,
  Phys.\ Rev.\ D {\bf 65}, 054506 (2002)
  [hep-lat/0110032];
  Phys.\ Rev.\ D {\bf 67}, 034501 (2003)
  [hep-lat/0211015];
  Phys.\ Rev.\ D {\bf 68}, 034506 (2003)
  [hep-lat/0303013].

\bibitem{Taylor} 
  J.~C.~Taylor and S.~M.~H.~Wong,
  Nucl.\ Phys.\ B {\bf 346}, 115 (1990);
  E.~Braaten and R.~D.~Pisarski,
  Phys.\ Rev.\ D {\bf 45}, 1827 (1992).

\bibitem{Ipp} 
  A.~Ipp, K.~Kajantie, A.~Rebhan and A.~Vuorinen,
  Phys.\ Rev.\ D {\bf 74}, 045016 (2006)
  [hep-ph/0604060].

\bibitem{KLRS2}
K.~Kajantie, M.~Laine, K.~Rummukainen and Y.~Schr\"oder,
JHEP {\bf 0304}, 036 (2003)
[hep-ph/0304048].

\bibitem{1lbe}
  D.~J.~Gross and F.~Wilczek,
  Phys.\ Rev.\ Lett.\  {\bf 30}, 1343 (1973);
  H.~D.~Politzer,
  Phys.\ Rev.\ Lett.\  {\bf 30}, 1346 (1973).

\bibitem{2lbe}
  W.~E.~Caswell,
  Phys.\ Rev.\ Lett.\  {\bf 33}, 244 (1974);
  D.~R.~T.~Jones,
  Nucl.\ Phys.\ B {\bf 75}, 531 (1974);
  E.~Egorian and O.~V.~Tarasov,
  Theor.\ Math.\ Phys.\  {\bf 41}, 863 (1979)
  [Teor.\ Mat.\ Fiz.\  {\bf 41}, 26 (1979)].

\bibitem{'tHooft:1973mm}
  G.~'t Hooft,
  Nucl.\ Phys.\ B {\bf 61}, 455 (1973).
 
\bibitem{3lbebMS}
  O.~V.~Tarasov, A.~A.~Vladimirov and A.~Y.~Zharkov,
  Phys.\ Lett.\ B {\bf 93} (1980) 429;
  S.~A.~Larin and J.~A.~M.~Vermaseren,
  Phys.\ Lett.\ B {\bf 303}, 334 (1993)
  [hep-ph/9302208].
 
\bibitem{4lbebMS}
  T.~van Ritbergen, J.~A.~M.~Vermaseren and S.~A.~Larin,
  Phys.\ Lett.\ B {\bf 400}, 379 (1997)
  [hep-ph/9701390].

\bibitem{GCCV}
  G.~Cveti\v{c} and C.~Valenzuela,
  Phys.\ Rev.\  D {\bf 74}, 114030 (2006)
  [hep-ph/0608256];
  J.\ Phys.\ G {\bf 32}, L27 (2006)
  [hep-ph/0601050].

\end{thebibliography}
\end{document}